\documentclass[lettersize,journal]{IEEEtran}
\usepackage{amsmath,amsfonts,amssymb}
\usepackage{algorithm}
\usepackage{algpseudocode}
\algrenewcommand\algorithmicrequire{\textbf{Input:}}
\algrenewcommand\algorithmicensure{\textbf{Output:}}
\usepackage{array}
\usepackage{booktabs}
\usepackage{multirow}
\usepackage{textcomp}
\usepackage{stfloats}
\usepackage{url}
\usepackage{verbatim}
\usepackage{graphicx}
\usepackage{xcolor}
\usepackage{cite}
\makeatletter
\providecommand\@LN[2]{}
\providecommand\@LN@col[1]{}
\makeatother

\begin{document}

\title{Empirically Calibrated Differential Games for Simulating Heterogeneous Driver Interactions at Urban Intersections}

\author{Amirmohammad~Khakpour,
        Nachuan~Li,
        Pedram~Beigi,
        and~Yu~(Marco)~Nie%
\thanks{A. Khakpour, N. Li, and Y. (M.) Nie are with the Department of Civil and Environmental Engineering, Northwestern University, Evanston, IL 60208 USA (e-mail: Amirmohammadkhakpour2029@u.northwestern.edu; Nachuan.li@northwestern.edu; y-nie@northwestern.edu). \textit{(Corresponding author: Yu (Marco) Nie.)}}%
\thanks{P. Beigi is with the Department of Civil and Environmental Engineering, George Washington University, Washington, DC 20052 USA (e-mail: beigi@gwu.edu).}}

\markboth{IEEE Transactions on Intelligent Transportation Systems}%
{Khakpour \MakeLowercase{\textit{et al.}}: Empirically Calibrated Differential Games for Simulating Driver Interactions}

\maketitle

\begin{abstract}
Game-theoretic models can resolve vehicle interactions without hand-written priority rules, but whether they describe human drivers once calibrated has rarely been tested. This study formulates a general-sum differential game of vehicle interactions at urban intersections, calibrates it to observed driving, and evaluates it on data not used for calibration. Each vehicle minimizes a cost that combines speed tracking, adherence to a map-derived lane route, control effort, gap keeping based on the intelligent driver model, a predictive conflict term, and a contact barrier, while perceiving the other vehicles with a delay. The game is solved with a feedback iterative linear--quadratic algorithm, and ten behavioral parameters per vehicle are calibrated by Bayesian optimization on trajectories recorded at four intersections in Washington, D.C. The calibrated game reproduces the trajectories of 819 vehicles with a mean position error of 1.00\,m. When only each vehicle's route is known, as in a simulator, it is the most accurate of the compared models on 96 held-out vehicles (7.42\,m), and ten-fold cross-validation shows that it is significantly more accurate than calibrated SUMO (by 3.61\,m), an inverse reinforcement learning model (by 0.93\,m), and independent and cooperative variants of the same model (by 0.33 and 0.32\,m). The solutions, computed in 2.1\,s for a typical interaction, are verified approximate equilibria. Finally, the conflict and steering weights, which the data do not identify, can be set from safety and comfort targets; together with a stiffer contact barrier, this removes the collisions that otherwise occur in 43\% of tight closed-loop conflicts.
\end{abstract}


\section{Introduction}
\IEEEPARstart{M}{icroscopic} traffic simulation is a standard tool for signal control, automated-vehicle (AV) testing, safety assessment, and street design \cite{hegde2025complete1,hegde2026complete2,hegde2026street}. Its predictive power rests on the behavioral model: how road users perceive one another, negotiate conflicts, and choose their motion. Even among human-driven vehicles (HDVs), drivers differ widely in reaction time and in their willingness to yield, which makes heterogeneous interactions a central challenge for simulation.

Existing simulators resolve interactions with rule-based car-following, lane-changing, and gap-acceptance models \cite{chandler_traffic_1958,gazis_car-following_1959,kesting2007general,kesting_enhanced_2010}. These rules work well in structured, homogeneous traffic. At urban intersections, however, road users with different objectives negotiate rights-of-way simultaneously, which requires ever more mode-specific rules and conflict heuristics, and the frameworks become hard to extend as the mix of street users diversifies \cite{ni2020limitations,zhao2022microscopic,anvari2015modelling,hegde2026street}.

Game theory replaces these rules with a single principle: each road user optimizes its own objective while anticipating the responses of the others. In traffic science, it was first applied to discrete decisions such as whether to merge, change lanes, or give way \cite{kita1999merging,talebpour2015modeling,barmpounakis2016modeling}. Several such models, calibrated on recorded trajectories, predict the observed decisions better than rule-based alternatives \cite{kang2017game,ali2019game}, and a review concludes that the potential of game theory for describing human drivers remains underused \cite{ji2020review}. Discrete games, however, decide only which vehicle goes first; the trajectories must still come from a separate car-following or path model.

Differential games remove this separation by letting the players choose their continuous motion \cite{isaacs1965differential,basar1999dynamic,bressan2011noncooperative}, so that the order of passage and the trajectories emerge together from one mechanism that accommodates heterogeneous objectives without ad hoc rules. Most of this work comes from automated driving, where games plan merging, overtaking, racing, and intersection crossing \cite{sadigh2016planning,fisac2019hierarchical,li2019differential,wang2021game,hang2022driving}, encode social preferences \cite{schwarting2019social}, and are solved for several vehicles in real time by iterative linear--quadratic and Newton-type methods \cite{fridovich2020efficient,laine2023computation}, sometimes blended with behavior learned from data \cite{lidard2024blending}. In these studies, however, the game plans the motion of an automated vehicle, the interaction is often reduced to a leader--follower, pairwise, or hierarchical structure \cite{li2020game,fang2023cooperative,davis2019merging}, although recorded interactions show both simultaneous and sequential responses \cite{li2026choose}, and the behavioral weights are set by hand rather than compared with human driving.

A growing body of work estimates the parameters of a game from observed trajectories. Inverse game methods recover the players' objectives from noisy and partial observations under open-loop \cite{peters2021inferring,peters2023online} and feedback \cite{li2023cost} Nash equilibria, and maximum-entropy formulations account for boundedly rational players \cite{mehr2023maximum}. Game-theoretic layers have been trained on recorded highway merges \cite{geiger2021learning}, and a quantal-response game of unprotected left turns has been calibrated on drone trajectories \cite{lian2025game}. These studies are mostly validated in simulation, on one maneuver, or on short-horizon prediction. They do not test whether a calibrated game, given only each vehicle's route as a simulator would be, reproduces human interactions better than the models that simulators use today, and they assume that drivers perceive each other without the delays known to shape car following \cite{treiber2006delays}.

Two other families of data-driven models are relevant. The first describes each driver as an optimal controller whose cost is learned by inverse reinforcement learning (IRL) \cite{ziebart2008maximum,sun2018probabilistic,wu2020efficient,huang2022driving} or calibrated in an optimal-control path model of intersections \cite{zhao_two-dimensional_2020,zhao2022microscopic}; the other vehicles enter as fixed predictions or through a shared utility, so no driver's plan depends on the self-interested responses of the others. The second family learns simulation agents from large driving datasets \cite{suo2021trafficsim,huang2023gameformer,montali2023waymo,wu2024smart}. These agents are realistic on public benchmarks, but their parameters carry no behavioral meaning that could be compared across drivers or set from safety targets. Whether a calibrated game predicts human interactions better than rule-based simulators and IRL driver models, and whether it does so because the drivers respond to each other, has not been established.

We address this question with the building block of a game-theoretic simulator: an $N$-player differential game that resolves one local interaction over a short horizon. The Nash equilibrium serves as a computational mechanism for conflict resolution, not as a claim that drivers solve games. Every vehicle is a player with its own behavioral parameters, calibrated from observed trajectories by Bayesian optimization (BO); the calibrated game is then evaluated on interactions not used for calibration and simulated in closed loop on conflicts tighter than any in the recordings. The contributions are:
\begin{enumerate}
\item \emph{Model and solver.} An $N$-player general-sum differential game of HDV interactions at intersections, with a map-derived lane route, an intelligent-driver-model (IDM) gap-keeping term, a predictive conflict term, a contact barrier, and delayed perception, solved by a feedback iterative linear--quadratic (iLQ) game algorithm \cite{fridovich2020efficient} adapted to nonconvex interaction costs and control bounds.
\item \emph{Calibration.} Per-vehicle Bayesian calibration of ten behavioral parameters, including the perception delay, on 819 vehicles of the TGSIM Foggy Bottom dataset, with nested-model tests showing that the interaction terms and the delay significantly improve the fit.
\item \emph{Evaluation.} A comparison under equal information with natively calibrated SUMO \cite{lopez2018microscopic}, sampling-based maximum-entropy IRL (SMIRL) \cite{wu2020efficient}, and independent and cooperative optimal control (OC), two variants of our model that change only how the vehicles are coupled. With only the route known, the game is the most accurate model on held-out interactions and in cross-validation; its solutions are verified approximate equilibria, computed faster than real time.
\item \emph{Identifiability and safety.} An analysis of which parameters the trajectories identify, with closed-loop simulations showing that two unidentified ones, the conflict and steering weights, govern collisions and turning comfort; set from safety and comfort targets and combined with a stiffer contact barrier, they lower the share of tight conflicts that end in a collision from 43\% to 0\%.
\end{enumerate}
Section~\ref{problem_setting} states the problem, Sections~\ref{sec:model} and~\ref{sec:solution_algorithm} formulate and solve the game, Section~\ref{sec:calibration} describes the calibration, Section~\ref{sec:experiments} the data and comparison models, Section~\ref{sec:results} the results, and Section~\ref{sec:conclusion} concludes.

\section{Problem setting}
\label{problem_setting}
Consider a simulator in which vehicles repeatedly meet and resolve local conflicts as they cross an urban intersection. Instead of priority and gap-acceptance rules, each local conflict is modeled as a finite-horizon differential game among the vehicles involved, and the equilibrium determines how the interaction unfolds. As vehicles enter and leave, conflicts appear and dissolve, so the simulator is a sequence of local games; this paper develops, calibrates, and evaluates the local game.

An interaction group $\mathcal{N}=\{1,\dots,N\}$ of HDVs plays over the horizon $[0,T]$; the formulation extends directly to AVs, pedestrians, and cyclists. The state of agent $i$ is $X_i(t)=[p_{xi}(t),\,p_{yi}(t),\,\theta_i(t),\,v_i(t)]^{\top}$, with position $(p_{xi},p_{yi})$, heading $\theta_i$, and speed $v_i$. Each agent enters the game with an observed state $X_i(0)$ and an intended destination $d_i$, and it chooses the controls $U_i(t)=[\kappa_i(t),\,a_i(t)]^\top$, its path curvature and longitudinal acceleration. The agents are mechanically independent, since each one's dynamics depend only on its own state and control, but strategically coupled: agent $i$ minimizes $J_i(U_i,U_{-i})$, which depends on the controls $U_{-i}$ of the others, perceived with a delay as $X_{-i}(t-\tau)$. The result is a general-sum game whose equilibrium controls $\{U_i^*\}$ and trajectories $\{X_i^*\}$ describe how every agent seeks efficient, comfortable, and safe passage while negotiating conflicts with the others.

\section{Model}
\label{sec:model}

Every vehicle is a player that balances four goals: (i)~travel at its preferred speed, (ii)~follow its lane through the intersection, (iii)~drive smoothly, and (iv)~avoid the others by keeping a safe gap to the vehicle ahead, steering clear of predicted conflicts, and never touching another vehicle. Drivers weigh these goals differently, so each vehicle has ten behavioral parameters estimated from data, while all other constants are shared (Table~\ref{tab:all_params}). Because each player's best plan depends on the others' plans, the plans are solved jointly as a Nash equilibrium.

\begin{table*}[!t]
\caption{Model Parameters: The Ten Behavioral Parameters Calibrated per Vehicle, and the Main Fixed Constants}
\label{tab:all_params}
\centering
\footnotesize
\begin{tabular}{@{}l l l l@{}}
\hline
Symbol & Meaning & Appears in & Search range / value\\
\hline
\multicolumn{4}{@{}l}{\emph{Calibrated per vehicle}}\\
$w_v$ & weight of tracking the reference speed & \eqref{eq:speed} & $[1,1000]$ (log scale)\\
$\bar v$ & desired cruise speed & \eqref{eq:ramp} & $[1,15]$ m/s\\
$w_\kappa$ & weight of lateral acceleration (steering effort) & \eqref{eq:control} & $[0.1,100]$ (log scale)\\
$w_a$ & weight of longitudinal acceleration & \eqref{eq:control} & $[1,500]$ (log scale)\\
$b_p$ & weight of reaching the destination position & \eqref{eq:terminal} & $[10,2000]$ (log scale)\\
$b_v$ & weight of reaching the terminal speed & \eqref{eq:terminal} & $[1,500]$ (log scale)\\
$b_\theta$ & weight of reaching the terminal heading & \eqref{eq:terminal} & $[1,500]$ (log scale)\\
$w_g$ & weight of keeping the IDM gap to the leader & \eqref{eq:follow} & $[10,1000]$ (log scale)\\
$w_c$ & weight of avoiding predicted conflicts & \eqref{eq:conflict} & $[10,1000]$ (log scale)\\
$\tau$ & perception delay toward interaction partners & \eqref{eq:delay} & $[0,3]$ s\\
\hline
\multicolumn{4}{@{}l}{\emph{Fixed for all vehicles}}\\
$a_r$ & rate of the speed-reference ramp & \eqref{eq:ramp} & 0.5 m/s$^2$\\
$r_{\rm stop}$, $\Delta_r$ & arrival distance and taper length of the speed reference & \eqref{eq:stopgate} & 1 m, 5 m\\
$w_{\rm lane}$ & weight of lane-route tracking & \eqref{eq:lane} & 50\\
$w_{\rm obs}$, $\epsilon_{\rm obs}$ & road-boundary repulsion weight and regularization & \eqref{eq:road} & 100, 0.1 m$^2$\\
$s_0$, $T_h$, $a_c$, $b_c$ & IDM jam gap, time headway, comfortable acceleration and deceleration & \eqref{eq:idm_gap} & 2 m, 1.9 s, 1.5 m/s$^2$, 2 m/s$^2$\\
$R_c$, $\mathcal{T}$ & conflict radius and look-ahead times & \eqref{eq:conflict} & 3 m, $\{0.5,1,1.5\}$ s\\
$W_c$ & contact-barrier weight & \eqref{eq:contact} & 1000 ($10^5$, rectangular barrier)\\
$\kappa_{\max}$, $a_{\min}$, $a_{\max}$ & control bounds & \eqref{eq:admissible} & 0.25 m$^{-1}$, $-6$ m/s$^2$, 3.5 m/s$^2$\\
$v_{\max}$, $a^{\rm lat}_{\max}$ & soft speed and lateral-acceleration limits & \eqref{eq:limits} & 15 m/s, 4 m/s$^2$\\
$\Delta t$ & time step & \eqref{eq:dynamic} & 0.1 s\\
\hline
\end{tabular}
\end{table*}

\subsection{Agent model}
\label{sec:agent_model}

Each agent $i\in \mathcal{N}$ minimizes a cost made of a terminal cost $\psi_i$ on its final state and a running cost $L_i$ accumulated over the horizon:
\begin{equation}
\label{eq:objective}
\begin{aligned}
J_i ={}&
\underbrace{\psi_i(X_i(T))}_{\text{Terminal Cost}}\\
&+
\underbrace{\int_{0}^{T} L_i\!\left(X_i(t),X_{-i}(t-\tau),U_i(t),t\right)dt}_{\text{Running Cost}}.
\end{aligned}
\end{equation}
The running cost depends on the agent's own state and control, on the delayed states of the others, and, through the speed reference, on time. Each agent follows the kinematic bicycle model \cite{polack2017kinematic} referenced at the rear axle. With the path curvature $\kappa_i=\tan\varphi_i/L$ (steering angle $\varphi_i$, wheelbase $L$) as input, this model is equivalent to a curvature-controlled unicycle:
\begin{equation}
\label{eq:dynamic}
\dot{X}_i(t)=f(X_i(t),U_i(t))
=
\begin{bmatrix}
v_i(t)\cos\theta_i(t) \\
v_i(t)\sin\theta_i(t) \\
v_i(t)\kappa_i(t) \\
a_i(t)
\end{bmatrix},
\end{equation}
discretized by forward Euler with $\Delta t=0.1$\,s, the sampling interval of the data. We write $p_i=(p_{x,i},p_{y,i})$ for the position, $e_i=(\cos\theta_i,\sin\theta_i)$ for the heading vector, and $n_i=(-\sin\theta_i,\cos\theta_i)$ for the left normal.

\subsubsection{Agent costs}
\label{sec:agent_costs}

\emph{Where the driver wants to be.} The terminal cost pulls the vehicle toward its destination:
\begin{equation}
\label{eq:terminal}
\begin{aligned}
\psi_i(X_i(T))
={}&
\frac{b_{p,i}}{2}\left[\left(p_{x,i}-p_{x,i}^{*}\right)^2+\left(p_{y,i}-p_{y,i}^{*}\right)^2\right]
\\
&+
\frac{b_{v,i}}{2}\left(v_i-v_i^{*}\right)^2
+
\frac{b_{\theta,i}}{2}
\left[
2\sin\!\left(
\frac{\theta_i-\theta_i^{*}}{2}
\right)
\right]^2,
\end{aligned}
\end{equation}
with destination $d_i=(p_{x,i}^{*},p_{y,i}^{*})$, terminal speed $v_i^{*}=0$, terminal heading $\theta_i^{*}$ (direction $e_i^{*}$), and calibrated weights $b_{p,i}$, $b_{v,i}$, $b_{\theta,i}$. The heading error is written in periodic form to avoid the discontinuity at $\pm\pi$.

\emph{How the driver gets there.} The running cost combines the driver's own goals with its interactions:
\begin{equation}
\label{eq:running}
\begin{aligned}
L_i
=&\;
\ell_i^{\text{speed}}
+
\ell_i^{\text{lane}}
+
\ell_i^{\text{control}}
+
\ell_i^{\text{road}}
+
\ell_i^{\text{rev}}
+
\ell_i^{\text{lim}}
\\
&+
\sum_{j\in\mathcal{F}_i}\ell_{ij}^{\text{follow}}
+
\sum_{j\in\mathcal{C}_i}\ell_{ij}^{\text{conflict}}
+
\sum_{j\neq i}\ell_{ij}^{\text{contact}},
\end{aligned}
\end{equation}
where $\mathcal{F}_i$ contains the same-stream leader of agent $i$ (if any) and $\mathcal{C}_i$ its crossing-conflict partners. The own-goal terms are:

\begin{figure}[!t]
\centering
\includegraphics[width=0.95\columnwidth]{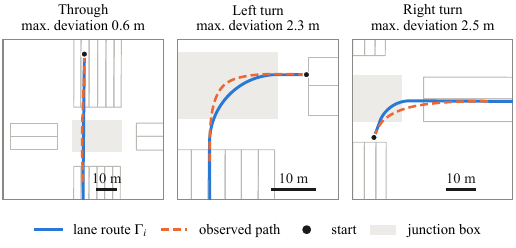}
\caption{Map-derived lane routes $\Gamma_i$ and observed paths of a through, a left-turning, and a right-turning agent, each with the median maximum deviation of its movement.}
\label{fig:lane_routes}
\end{figure}

\begin{description}
\item[1.] $\ell_i^{\text{speed}}$ draws the vehicle's speed toward a reference that approaches the preferred speed gradually:
\begin{equation}
\label{eq:speed}
\ell_i^{\text{speed}}
=
\frac{w_{v,i}}{2}
\left(
v_i(t)-g_i(p_i)\,v_i^{\text{ref}}(t)
\right)^2,
\end{equation}
where the reference ramps from the initial speed $v_{i,0}$ toward the desired speed $\bar v_i$ at a comfortable rate $a_r$:
\begin{equation}
\label{eq:ramp}
v_i^{\text{ref}}(t)=v_{i,0}+\operatorname{sgn}(\bar v_i-v_{i,0})\min\!\left(|\bar v_i-v_{i,0}|,\;a_r t\right),
\end{equation}
and the gate $g_i$ tapers it to zero on the final approach:
\begin{equation}
\label{eq:stopgate}
g_i(p_i)=\mathcal{S}\!\left(\frac{(d_i-p_i)^{\top}e_i^{*}-r_{\text{stop}}}{\Delta_r}\right),
\end{equation}
with $(d_i-p_i)^{\top}e_i^{*}$ the remaining distance to the destination and $\mathcal{S}$ the smoothstep of Section~\ref{sec:smoothing}. Unlike a constant reference, the ramp avoids abrupt starts.

\item[2.] $\ell_i^{\text{lane}}$ keeps the vehicle on a lane route $\Gamma_i$ built from the intersection map:
\begin{equation}
\label{eq:lane}
\ell_i^{\text{lane}}
=
\frac{w_{\text{lane}}}{2}\,e_{\Gamma_i}(p_i(t))^2,
\end{equation}
where $e_{\Gamma_i}(p)$ is the signed lateral distance to $\Gamma_i$. The route is a straight approach along the approach lane, a cubic B\'ezier turn \cite{gonzalez2016review}, and a straight exit along the exit lane; the straight parts pass through the vehicle's own start and end points, so the map decides only where the turn takes place, and turning vehicles do not cut corners as they would under a direct attraction to the destination. Observed paths deviate a median 0.30\,m from their routes, mostly in turns (Fig.~\ref{fig:lane_routes}).

\item[3.] $\ell_i^{\text{control}}$ penalizes steering and acceleration effort:
\begin{equation}
\label{eq:control}
\begin{aligned}
\ell_i^{\text{control}}
={}&
\frac{w_{\kappa,i}}{2}\kappa_i^2(t)v_i^4(t)
+
\frac{w_{a,i}}{2}a_i^2(t)
\\
&+\frac{\epsilon_\kappa}{2}\kappa_i^2(t)
+\frac{w_{\text{low}}}{2}\left(1-\mathcal{S}\!\left(\frac{v_i(t)}{v_{\text{low}}}\right)\right)\kappa_i^2(t).
\end{aligned}
\end{equation}
The first term is the squared lateral acceleration $(v_i^2\kappa_i)^2$, and the second is the squared longitudinal acceleration. The last two terms ($\epsilon_\kappa=10^{-3}$, $w_{\text{low}}=5$, $v_{\text{low}}=1$\,m/s) keep the curvature well conditioned near standstill.

\item[4.] $\ell_i^{\text{road}}$ repels the vehicle from road boundaries:
\begin{equation}
\label{eq:road}
\ell_i^{\text{road}}
=
\sum_{o_m\in\mathcal{O}}
\frac{w_{\text{obs}}}
{\left\|p_i(t)-o_m\right\|^2+\epsilon_{\text{obs}}},
\end{equation}
where $\mathcal{O}$ is the set of boundary points. This is an artificial potential field \cite{khatib1986real,wolf2008artificial}; because the repulsion decays with the squared distance, it acts only close to the road edge.

\item[5.] $\ell_i^{\text{rev}}$ prevents reversing:
\begin{equation}
\label{eq:reverse}
\ell_i^{\text{rev}}=\frac{w_{\text{rev}}}{2}\,\mathcal{P}_{\sigma}\!\left(-v_i(t)\right)^2,
\end{equation}
with the softplus $\mathcal{P}_{\sigma}$, $w_{\text{rev}}=50$, and $\sigma=0.1$\,m/s.

\item[6.] $\ell_i^{\text{lim}}$ enforces the speed and lateral-acceleration limits softly:
\begin{equation}
\label{eq:limits}
\ell_i^{\text{lim}}=h\!\left(v_i-v_{\max}\right)+h\!\left(v_{\min}-v_i\right)+h\!\left(|v_i^2\kappa_i|-a^{\text{lat}}_{\max}\right),
\end{equation}
with the $C^2$ hinge $h$, $[v_{\min},v_{\max}]=[0,15]$\,m/s, and $a^{\text{lat}}_{\max}=4$\,m/s$^2$.
\end{description}

\emph{How the driver interacts.} Drivers keep a gap to the vehicle ahead, avoid crossing traffic, and never make contact with another vehicle; the partners are inferred from the scene geometry. Two agents share a stream if their initial headings differ by less than $30^\circ$ and their lateral offset is at most 3.5\,m; the nearest same-stream agent ahead is the \emph{leader} ($\mathcal{F}_i$), and every other-stream agent is a \emph{conflict partner} ($\mathcal{C}_i$). Below, $\delta_{ij}=p_j-p_i$ and $v^{+}=\mathcal{P}_{0.1}(v)$.

\begin{description}
\item[7.] $\ell_{ij}^{\text{follow}}$ keeps the intelligent-driver-model (IDM) gap \cite{treiber2000congested} to the leader. With vehicle length $L_v=4.5$\,m, let
\begin{equation}
s_{ij}=\delta_{ij}^{\top}e_i-L_v,\qquad
l_{ij}=\delta_{ij}^{\top}n_i,\qquad
\Delta v_{ij}=v_i^{+}-v_j^{+}\,(e_i^{\top}e_j)
\end{equation}
be the bumper gap, lateral offset, and closing speed. The desired gap is
\begin{equation}
\label{eq:idm_gap}
s_{ij}^{\star}=s_0+\mathcal{P}_{0.3}\!\left(v_i^{+}T_h+\frac{v_i^{+}\Delta v_{ij}}{2\sqrt{a_c b_c}}\right),
\end{equation}
with the standard IDM constants $s_0=2$\,m, $a_c=1.5$\,m/s$^2$, $b_c=2$\,m/s$^2$, and the headway $T_h=1.9$\,s measured in the dataset. Only gaps shorter than desired are penalized:
\begin{equation}
\label{eq:follow}
\ell_{ij}^{\text{follow}}
=
\frac{w_{g,i}}{2}\,G_{\text{lat}}(l_{ij})
\left[\frac{\mathcal{P}_{0.3}\!\left(s_{ij}^{\star}-s_{ij}\right)}{5\ \text{m}}\right]^2,
\end{equation}
where $G_{\text{lat}}(l)=\exp\!\left(-l^2/(2\cdot 2.5^2)\right)$ fades the term as the leader leaves the follower's corridor. The IDM thus enters the game as a cost term rather than replacing the vehicle dynamics.

\item[8.] $\ell_{ij}^{\text{conflict}}$ avoids predicted close approaches with crossing traffic. With relative velocity $u_{ij}=v_j^{+}e_j-v_i^{+}e_i$ and look-ahead times $\sigma\in\mathcal{T}=\{0.5,\,1.0,\,1.5\}$\,s,
\begin{equation}
\label{eq:conflict}
\ell_{ij}^{\text{conflict}}
=
\frac{w_{c,i}}{6}
\sum_{\sigma\in\mathcal{T}}
\exp\!\left(-\frac{\left\|\delta_{ij}+\sigma\,u_{ij}\right\|^2}{2R_c^2}\right),
\end{equation}
with $R_c=3$\,m. The term is a smooth surrogate for the event that both vehicles occupy the same space within the next 1.5\,s, in the spirit of potential-field collision costs \cite{wolf2008artificial} evaluated at constant-velocity predictions. Unlike time to collision, it never divides by small speeds.

\item[9.] $\ell_{ij}^{\text{contact}}$ is a barrier on an elliptical outline against every other agent:
\begin{equation}
\label{eq:contact}
\begin{aligned}
\rho_{ij}&=\sqrt{\left(\frac{\delta_{ij}^{\top}e_i}{4.5}\right)^2+\left(\frac{\delta_{ij}^{\top}n_i}{1.8}\right)^2+0.01},\\
\ell_{ij}^{\text{contact}}&=\frac{W_c}{2}\,\mathcal{P}_{0.08}\!\left(1-\rho_{ij}\right)^2,
\end{aligned}
\end{equation}
with $W_c=1000$, which prevents two players from passing through each other. The ellipse is the default used throughout calibration and evaluation. For safety-critical simulation it can optionally be replaced by a \emph{rectangular barrier} built from the two vehicles' outlines: with relative heading $\phi_{ij}=\theta_j-\theta_i$, recorded lengths $L_i$, $L_j$, width $W=1.8$\,m, and margin $m=0.25$\,m,
\begin{equation}
\label{eq:contact-box}
\begin{aligned}
a_{ij}&=\tfrac12\big(L_i+L_j|\cos\phi_{ij}|+W|\sin\phi_{ij}|\big)+m,\\
b_{ij}&=\tfrac12\big(W+L_j|\sin\phi_{ij}|+W|\cos\phi_{ij}|\big)+m,\\
\rho_{ij}&=\left[\left(\frac{\delta_{ij}^{\top}e_i}{a_{ij}}\right)^{16}+\left(\frac{\delta_{ij}^{\top}n_i}{b_{ij}}\right)^{16}\right]^{1/16},
\end{aligned}
\end{equation}
a smoothed box that contains every overlap of the two rectangles and is inactive until the vehicles are close to contact. Used with $W_c=10^5$ (Section~\ref{sec:sensitivity}), it is much stiffer and needs more solver iterations, so it is an option for simulation rather than part of the base model.
\end{description}

\emph{What the driver perceives.} Drivers react to what they saw a moment ago \cite{treiber2006delays}, so the following and conflict terms use the partner's \emph{delayed} state. With $D_i=\mathrm{round}(\tau_i/\Delta t)$, agent $i$ perceives at step $k$
\begin{equation}
\label{eq:delay}
\widetilde{X}_j(k)=
\begin{cases}
X_j(k-D_i), & k\ge D_i,\\
X_j^{\text{obs}}(k-D_i), & k<D_i,
\end{cases}
\end{equation}
where $X_j^{\text{obs}}$ is the partner's state before the game starts (extrapolated backward at constant velocity where not recorded). The agent's own state and the contact barrier are not delayed.

\subsubsection{Constraints}
\label{sec:constraints}
Each agent starts from its measured state, $X_i(0)=X_{i,0}$, and its controls lie in the admissible set
\begin{equation}
\label{eq:admissible}
\mathcal{U}_i=\left\{(\kappa_i,a_i):\kappa_{\min}\le \kappa_i \le \kappa_{\max},\; a_{\min}\le a_i \le a_{\max}\right\},
\end{equation}
with $\kappa_{\max}=-\kappa_{\min}=0.25$\,m$^{-1}$, $a_{\min}=-6$\,m/s$^2$, and $a_{\max}=3.5$\,m/s$^2$, which contain all observed kinematics. State limits are enforced softly by $\ell_i^{\text{rev}}$ and $\ell_i^{\text{lim}}$.

\subsubsection{Smooth approximations}
\label{sec:smoothing}
The solver needs the gradient and Hessian of every cost, so each nonsmooth operator is replaced by a smooth one, with the widths given above: the \emph{softplus} $\mathcal{P}_w(z)=w\log(1+e^{z/w})$, a $C^\infty$ approximation of $\max(z,0)$; the \emph{quintic smoothstep} $\mathcal{S}(s)=6s^5-15s^4+10s^3$ on $[0,1]$ \cite{perlin2002improving}, equal to 0 below and 1 above that interval, with first and second derivatives vanishing at both ends; the \emph{$C^2$ quadratic hinge} $h(z)$, which is $0$ for $z\le0$, $W\varepsilon^2(\tfrac32 s^3-\tfrac32 s^4+\tfrac12 s^5)$ with $s=z/\varepsilon$ for $0<z<\varepsilon$, and $\tfrac12 Wz^2$ for $z\ge\varepsilon$ ($W=1000$, $\varepsilon=0.1$); \emph{Gaussian kernels} $\exp(-\|\cdot\|^2/2R^2)$ in place of the indicators ``same lane'' and ``on collision course''; and the \emph{periodic heading error} $2\sin(\Delta\theta/2)$ in place of angle wrapping. The lane cost and items 5--9 penalize a residual $r(x)$ through a convex function $\phi$, either $\tfrac12 r^2$ or $h(r)$. The exact Hessian of such a term, $\phi''(r)\,\nabla r\,\nabla r^{\top}+\phi'(r)\,\nabla^2 r$, can be indefinite because of its second part. The solver keeps only the positive-semidefinite first part (Gauss--Newton \cite{nocedal2006numerical}) and the exact gradient $\phi'(r)\nabla r$.

\subsubsection{Agent's best response}
\label{sec:agent_best_response}
Each driver chooses the plan that minimizes its own cost given the others' plans $U_{-i}$, its best response
\begin{equation}
U_i^{\mathrm{BR}}(U_{-i})
=
\operatorname*{arg\,min}_{U_i(\cdot)\in \mathcal{U}_i}
J_i(U_i,U_{-i}),
\end{equation}
an optimal-control problem \cite{bryson1975applied} subject to \eqref{eq:dynamic}, in which the delayed partner states are given data.

\subsection{Nash equilibrium}
\label{sec:nash_equilibrium}
When every driver's plan is a best response to the others', no driver can gain by deviating alone, and the plans $U^{*}=\{U_1^{*},\ldots,U_N^{*}\}$ form a Nash equilibrium \cite{basar1999dynamic}, that is, for every agent $i$,
\begin{equation}
J_i\!\left(U_i^{*},U_{-i}^{*}\right)
\le
J_i\!\left(U_i,U_{-i}^{*}\right),
\qquad
\forall U_i\in\mathcal{U}_i,
\end{equation}
with $\mathcal{U}_i$ from \eqref{eq:admissible}. The solver represents each strategy as an affine state feedback around the nominal trajectories (Section~\ref{sec:open_loop_nash}), which produces the controls $U_i^{*}$ along the equilibrium trajectories.

A nonlinear game can have several equilibria, corresponding to different passing orders or yielding decisions, and the one reached depends on the warm start and the parameters. Since the smoothed costs are twice differentiable and the dynamics smooth, local equilibria are well defined, and a local equilibrium is all a simulator needs to resolve each short-horizon conflict.

\section{Solution algorithm}
\label{sec:solution_algorithm}
We compute the equilibrium with the iterative linear--quadratic (iLQ) game method \cite{fridovich2020efficient} (Algorithm~\ref{alg:overall_framework}). Each iteration rolls every vehicle forward along its current plan, approximates the game around these trajectories by a linear--quadratic (LQ) game, solves it for every vehicle's \emph{feedback} strategy, and applies the correction with a step size chosen by a line search. Because all vehicles are corrected jointly, each plan accounts for how the others respond; this joint correction distinguishes the game from separate single-vehicle optimizations.

\begin{algorithm}[!t]
\footnotesize
\caption{Overall iterative solution framework}
\label{alg:overall_framework}
\textbf{Input:} initial state $X_0$, nominal controls $\bar U^{(0)}$, horizon $K=T/\Delta t$, tolerances $\epsilon_X$, $\epsilon_J$.\\
\textbf{Output:} candidate local Nash equilibrium $(X^*,U^*)$.
\begin{enumerate}
\item For $n=0,1,\ldots,n_{\max}$:
\begin{enumerate}
\item Roll out the nonlinear dynamics under $\bar U^{(n)}$ to obtain $\bar X^{(n)}$.
\item Linearize the dynamics and quadratize every agent's smoothed cost (Section~\ref{sec:smoothing}) along $(\bar X^{(n)},\bar U^{(n)})$, with Gauss--Newton curvature for residual-type costs and diagonal regularization.
\item Solve the local feedback LQ Nash game (Algorithm~\ref{alg:open_loop_nash}) for the gains $P_{i,k}$ and feedforward terms $\alpha_{i,k}$.
\item Choose $\eta$ by a safeguarded line search and update
$\bar U_{i,k}^{(n+1)}=\Pi_{\mathcal U}\big(\bar U_{i,k}^{(n)}-P_{i,k}(X_k-\bar X_k^{(n)})-\eta\,\alpha_{i,k}\big)$, where $\Pi_{\mathcal U}$ projects onto the admissible controls.
\item Stop if the state trajectories and costs meet the tolerances.
\end{enumerate}
\item Return the final iterate.
\end{enumerate}
\end{algorithm}
\subsection{Solution of the local feedback LQ Nash game}
\label{sec:open_loop_nash}

In terms of the deviations $\delta X_k=X_k-\bar X_k$ and $\delta U_{i,k}=U_{i,k}-\bar U_{i,k}$, the linearized dynamics are $\delta X_{k+1}=A_k\delta X_k+\sum_i B_{i,k}\delta U_{i,k}$. Agent $i$'s quadratic cost approximation consists of the state and control Hessians $Q_{i,k}$, $R_{ij,k}$, the gradients $l_{i,k}$, $r_{ij,k}$, and the mixed blocks $N_{ij,k}$, which arise from the speed--curvature coupling in \eqref{eq:control} and \eqref{eq:limits}. Each agent adopts an affine feedback strategy
\begin{equation}
\label{eq:strategy}
\delta U_{i,k}=-P_{i,k}\,\delta X_k-\alpha_{i,k},
\end{equation}
and its quadratic value function at step $k$ is described by a matrix $Z_{i,k}$ and a vector $\zeta_{i,k}$. Stationarity of every agent's stage problem, given the other agents' strategies, yields the coupled linear system
\begin{equation}
\label{eq:coupled}
\Lambda_k\,\big[\,P_k\;\;\alpha_k\,\big]=\big[\,Y_k\;\;y_k\,\big],
\end{equation}
where $P_k$ and $\alpha_k$ stack all agents' gains and feedforward terms. The block $(i,j)$ of the \emph{coupling matrix} $\Lambda_k$ equals $R_{ii,k}+B_{i,k}^{\top}Z_{i,k+1}B_{i,k}$ for $j=i$ and $B_{i,k}^{\top}Z_{i,k+1}B_{j,k}$ for $j\neq i$. The right-hand-side blocks are $Y_{i,k}=B_{i,k}^{\top}Z_{i,k+1}A_k+N_{ii,k}$ and $y_{i,k}=B_{i,k}^{\top}\zeta_{i,k+1}+r_{ii,k}$. With the closed-loop matrix $F_k=A_k-\sum_jB_{j,k}P_{j,k}$ and drift $\beta_k=-\sum_jB_{j,k}\alpha_{j,k}$, the value functions are propagated backward as
\begin{equation}
\label{eq:riccati}
\begin{aligned}
Z_{i,k}={}&F_k^{\top}Z_{i,k+1}F_k+Q_{i,k}+\sum_j P_{j,k}^{\top}R_{ij,k}P_{j,k}\\
&-\sum_j\big(N_{ij,k}^{\top}P_{j,k}+P_{j,k}^{\top}N_{ij,k}\big),\\
\zeta_{i,k}={}&F_k^{\top}\big(\zeta_{i,k+1}+Z_{i,k+1}\beta_k\big)+l_{i,k}\\
&+\sum_j P_{j,k}^{\top}\big(R_{ij,k}\alpha_{j,k}-r_{ij,k}\big)-\sum_jN_{ij,k}^{\top}\alpha_{j,k}.
\end{aligned}
\end{equation}
Algorithm~\ref{alg:open_loop_nash} summarizes the recursion.

\begin{algorithm}[!t]
\footnotesize
\caption{Solution of the local feedback LQ Nash game}
\label{alg:open_loop_nash}
\textbf{Input:}
Linearized dynamics $(A_k,B_{i,k})$, quadratic cost approximation $(Q_{i,k},l_{i,k},R_{ij,k},r_{ij,k},N_{ij,k})$, nominal controls $\bar U_k$, and bounds $\mathcal{U}$.

\textbf{Output:}
Feedback gains $\{P_{i,k}\}$ and feedforward terms $\{\alpha_{i,k}\}$.
\begin{enumerate}
\item Initialize the value functions at $k=K$ from the terminal cost.
\item \textbf{Backward recursion}
\begin{enumerate}
\item For $k=K-1,\ldots,0$:
\begin{enumerate}
\item Verify that each state cost matrix $Q_{i,k}$ is positive semi-definite and that each own-control block of $\Lambda_k$ is positive definite.
\item Construct the coupled matrix $\Lambda_k$ and right-hand sides of \eqref{eq:coupled}.
\item Solve \eqref{eq:coupled} for $(P_k,\alpha_k)$, holding at their bounds the controls whose nominal value plus step would leave $\mathcal{U}$.
\item If $\Lambda_k$ is ill-conditioned, increase the regularization and recompute.
\item Update each agent's value function by \eqref{eq:riccati}.
\end{enumerate}
\item End For.
\end{enumerate}
\item Return $\{P_{i,k},\alpha_{i,k}\}$. The forward pass is the nonlinear rollout in step~1(d) of Algorithm~\ref{alg:overall_framework}.
\end{enumerate}
\end{algorithm}
This is the game counterpart of the Riccati recursion: each agent's cost-to-go is propagated backward, but all agents' stationarity conditions are solved together through \eqref{eq:coupled}, with $\Lambda_k$ factorized once per step. A control that would leave its bounds is clamped and removed from the free subsystem \cite{tassa2014control}, so the search direction respects the bounds. The feedback law \eqref{eq:strategy} acts on the \emph{actual} deviation of the nonlinear rollout, which keeps each agent's response consistent with the others' updated motion. The LQ game has a unique feedback Nash solution when the state costs are positive semidefinite, the own-control blocks positive definite, and $\Lambda_k$ invertible \cite{basar1999dynamic}. Because the delayed partner state \eqref{eq:delay} is given data, the blocks that involve it are removed from agent $i$'s local model. In the resulting local equilibrium, each agent responds optimally to its delayed perception of the others.

\subsection{Numerical stabilization and convergence}
\label{sec:numerical_stability}
Because the interaction, contact, and limit costs are nonconvex, we use their Gauss--Newton curvature (Section~\ref{sec:smoothing}) and keep only the positive-semidefinite part of the speed--curvature cross terms. In addition, we apply the diagonal regularization $\widehat Q_{i,k}\leftarrow\widehat Q_{i,k}+\epsilon_x I$, $R_{ii,k}\leftarrow R_{ii,k}+\mu I$ with $\epsilon_x=10^{-6}$ and an adaptive Levenberg--Marquardt term $\mu$ \cite{tassa2012synthesis}.

The strategies are applied through a backtracking line search on the feedforward scale $\eta\in\{0.5,0.25,\ldots,1/32\}$. It accepts the largest step whose rollout is finite, does not increase the summed cost $\sum_iJ_i$ (used only as a merit function), and does not increase the largest state update more than fivefold. If no step is accepted, $\mu$ is increased tenfold (within $[10^{-3},10^{6}]$) and the LQ game is solved again. Convergence is declared when
\begin{gather}
\max_k\left\|X_k^{(n+1)}-X_k^{(n)}\right\|_\infty<\epsilon_X,\\
\frac{\left|\sum_iJ_i^{(n+1)}-\sum_iJ_i^{(n)}\right|}{\left|\sum_iJ_i^{(n)}\right|}<\epsilon_J
\end{gather}
hold for three consecutive iterations, with $\epsilon_X=10^{-2}$ and $\epsilon_J=10^{-6}$. A run that does not converge ends in one of three ways: it reaches the iteration \emph{budget}, it \emph{stalls} because no descent step is found (the best iterate is then kept), or it \emph{diverges}. Whether a solution is an equilibrium is verified separately (Section~\ref{sec:res_gap}).

\section{Calibration}
\label{sec:calibration}
The behavioral parameters are estimated so that the equilibrium reproduces the observed interaction. In this bilevel problem, the inner level solves the game for given parameters and the outer level adjusts the parameters to match the recorded trajectories. Because every outer evaluation requires solving a nonlinear game, the outer level uses Gaussian-process BO \cite{snoek2012practical}, which needs relatively few evaluations.
 Simulated and observed positions are compared at the same instants, so the model must reproduce timing as well as path. For the steps $k=0,\dots,K_i$ in which agent $i$ is present, the objective is the time-aligned root-mean-square (RMS) distance
\begin{equation}
E_i
=
\sqrt{
\frac{1}{K_i+1}
\sum_{k=0}^{K_i}
\left\|
\mathbf p_i^{\rm sim}(k\Delta t)
-
\mathbf p_i^{\rm obs}(k\Delta t)
\right\|_2^2
}.
\label{eq:rms_error}
\end{equation}
A vehicle that follows the right path but arrives one second late is thus penalized by roughly the distance it travels in one second. The objective of a case is the mean of $E_i$ over its agents, and the same error measure scores every model in Section~\ref{sec:results}. The ten parameters of Table~\ref{tab:all_params} are calibrated per agent, the weights on a $\log_{10}$ scale. The interaction parameters ($w_g$, $w_c$, $\tau$) act only when a partner is nearby, so they are calibrated for agents with a partner within 20\,m and held at their defaults otherwise.

\subsection{Bayesian calibration procedure}
Each case is built around one \emph{ego} vehicle, the vehicle whose movement through the intersection defines the case. The game window is anchored to the ego's passage through the \emph{junction box}, the intersection area that the TGSIM dataset delineates up to the crosswalks, excluding the crosswalks and sidewalks. The window starts 3\,s before the ego enters the box and ends 2\,s after it exits. Windows are limited to 10\,s by shortening the margins to as little as 1\,s, unless the passage alone exceeds 8\,s (31 of 412 cases). The players are the ego and up to five other vehicles, chosen nearest first, that are recorded throughout the window, travel at least 2\,m, and pass within 35\,m of the ego's path; vehicles present for only part of the window are not included. The players' initial states, destinations, and terminal headings come from the recording, and the road boundaries and lane routes from the map.

To avoid optimizing all agents' parameters at once, we use block coordinate descent: BO optimizes one agent's parameters with the others held fixed, every candidate is scored by re-solving the full game, and an update is accepted only if it lowers the case error. BO uses a Mat\'ern-$5/2$ Gaussian process and expected improvement \cite{jones1998efficient}, with 8 space-filling and 20 guided evaluations per agent in each sweep, and each game is solved with up to 50 iterations. Two sweeps are run, so every agent is revisited after its partners have been updated. Each solve starts from a warm start in which every agent drives toward its destination at its candidate desired speed while ignoring the others, so the interactions arise from the game itself. All seeds are fixed.

\subsection{Statistical significance of the behavioral mechanisms}
\label{sec:significance_method}
Per-parameter $t$-statistics are uninformative here, because residuals sampled at 10\,Hz are almost perfectly autocorrelated (median lag-1 autocorrelation 0.96). We therefore test each mechanism with a nested-model comparison, calibrating the model with it (\emph{full}) and without it (\emph{restricted}) on the same cases, budget, and seeds:
\begin{itemize}
\item \emph{Interaction terms}: the full model versus a model without the following, conflict, and contact terms (and hence without $w_g$, $w_c$, and $\tau$);
\item \emph{Perception delay}: the full model versus the same model with $\tau=0$.
\end{itemize}
A positive mean of $d_i=E_i^{\text{restricted}}-E_i^{\text{full}}$ indicates that the mechanism reduces the error. We test $t=\bar d/\mathrm{SE}(\bar d)$ with standard errors clustered by case \cite{cameron2015practitioner} and (number of cases $-$ 1) degrees of freedom. The test is applied to the agents with an interaction partner, and to all agents as a robustness check, both in-sample and on the held-out cases.

\section{Experiment setup}
\label{sec:experiments}
\subsection{Data}
\label{sec:data}

\begin{figure}[!t]
\centering
\includegraphics[width=0.85\columnwidth]{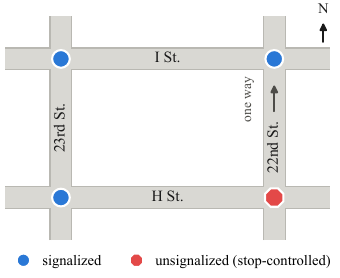}
\caption{Foggy Bottom study area (schematic) and the four analyzed intersections.}
\label{fig:site}
\end{figure}

We use the TGSIM Foggy Bottom dataset, recorded by infrastructure videography in Washington, D.C., in the evening rush hour at 10\,Hz \cite{ammourah2025introduction}. It covers four intersections (Fig.~\ref{fig:site}): I~St.\ \& 23~St., I~St.\ \& 22~St., and H~St.\ \& 23~St.\ are signalized, and H~St.\ \& 22~St.\ is not. Signal phases are not recorded, so all models are given the movement permissions observed in the recordings.

A \emph{case} is a group of vehicles that are present at the same time and interact within a common time window, and each vehicle in a case is an \emph{agent}, that is, a player of the game. Of the 416 extracted cases, 412 yield a window anchored to the junction box (median length 7.9\,s). The \emph{training} set has 370 cases with 819 agents: 703 signalized (348 left, 25 right, 326 through, 4 U-turns) and 116 unsignalized (54 left, 32 right, 30 through). The remaining 42 cases with 96 agents form the \emph{held-out} set.

\subsection{SUMO setup}
The rule-based reference is SUMO 1.23.1 \cite{lopez2018microscopic}, run natively on a network built from the same scene geometry, with the Krau{\ss} car-following model \cite{krauss1998microscopic} and SUMO's junction logic. Nine vehicle-type parameters (speedFactor, accel, decel, sigma, tau, minGap, actionStepLength, jmTimegapMinor, impatience) are calibrated per vehicle with the same cases, objective, BO budget, coordinate descent, and seeds as the game. Because SUMO's driver model is stochastic, every run is averaged over 10 random seeds.

\subsection{SMIRL setup}
The learning-based reference is the sampling-based maximum-entropy inverse reinforcement learning (SMIRL) driver model of Wu \emph{et al.} \cite{wu2020efficient}, which learns a linear reward over the original features (speed deviation, longitudinal and lateral acceleration, jerk, future distance, and future interaction distance) \cite{ziebart2008maximum} and predicts the minimum-cost sampled trajectory. Candidate paths follow the game's lane route with lateral offsets, with quartic speed profiles and stop--wait--go profiles for yielding; the interaction features are summed over the other agents, extrapolated at constant velocity from their initial states. The $\ell_1$ regularization strength is chosen by cross-validation. As in \cite{wu2020efficient}, one reward is learned for all drivers from the same 819 training vehicles.

\subsection{Independent and cooperative variants of the model}
\label{sec:oc_baselines}
Our model can couple the vehicles' problems in three ways. The Nash game of Section~\ref{sec:model} is the main formulation; the two alternatives below share its dynamics, cost terms, constraints, solver, warm start, and iteration budget and differ only in the coupling, so comparing them isolates the effect of solving a \emph{game}.

\emph{Independent optimal control (IOC).} Each vehicle minimizes its own cost $J_i$ against fixed predictions $\hat X_j$ of the others:
\begin{equation}
\label{eq:ioc}
U_i^{\mathrm{IOC}}=\operatorname*{arg\,min}_{U_i\in\mathcal{U}_i} J_i\big(U_i;\hat X_{-i}\big),
\end{equation}
where $\hat X_j$ is the warm-start trajectory of vehicle $j$. Each vehicle reacts to the others but assumes they do not react to it.

\emph{Cooperative OC.} All vehicles jointly minimize the team cost $\sum_{j\in\mathcal{N}}J_j$. The plans are coupled as in the game, but a vehicle accepts a higher cost of its own whenever this lowers the total, so the vehicles are not self-interested.

Both variants are calibrated separately with the same procedure as the game, so differences in accuracy reflect the coupling rather than parameters tuned for the game.

\subsection{Comparison setup}
\label{sec:benchmark}
The five models are compared on all 42 held-out cases (96 agents).

\paragraph*{Parameter transfer}
A simulator cannot calibrate road users it has never seen, so each held-out agent receives the mean calibrated parameters of its group, one of six combinations of signal control and movement (for SUMO the mean vehicle type, for the two variants their own group means, for SMIRL its single reward). Because a group-average delay does not transfer to new drivers (Section~\ref{sec:calibration_results}), the game predicts the held-out cases with $\tau=0$ and with the group means calibrated under that setting. Using the group-mean delay instead would raise its errors from 4.33 to 4.55\,m with the end location known and from 7.42 to 7.66\,m with it unknown.

\paragraph*{Information settings}
All models receive the same initial states, lane routes, and destination information, and none observes any position after $t=0$. In setting (1), the \emph{end location is known} but not its arrival time: the game and its variants place the terminal target there with a 15\,s horizon, SMIRL samples trajectories that end there at a free time, and SUMO stops the vehicle there. In setting (2), as in a simulator, the \emph{end location is unknown} and only the route is given: the terminal target lies where the vehicle's own speed reference \eqref{eq:ramp} would carry it. Errors are computed over each agent's observed window (mean 8.4\,s).

\paragraph*{Accuracy metrics}
For each agent we compute the time-aligned RMS errors of position \eqref{eq:rms_error}, speed, acceleration, and jerk, averaged with equal weight per agent; low acceleration and jerk errors indicate human-like smoothness. For comparison with trajectory forecasting, we also report the average displacement error $\mathrm{ADE}(h)$, the mean distance between predicted and observed positions over the first $h$ seconds, and the final displacement error $\mathrm{FDE}(h)$, the distance at time $h$; we also split the position error into along-track (timing) and lateral parts.

\paragraph*{Safety, convergence, and computation time}
A \emph{close approach} occurs when the centers of two vehicles come within 2\,m. The optimization-based models use the same tolerances and a budget of 100 iterations. Computation time is the single-thread wall time to produce all trajectories of a case (for SUMO, one run without network construction), and the \emph{real-time factor} divides it by the simulated horizon.

\paragraph*{Equilibrium verification}
For each player of every held-out solution, the others' controls are held fixed and $J_i$ is minimized over its own controls using L-BFGS-B \cite{byrd1995limited} with exact gradients, started from the solution and from two perturbations of it. The \emph{relative Nash gap} is
\begin{equation}
\label{eq:nashgap}
g_i=\frac{J_i\big(U_i^{*},U_{-i}^{*}\big)-\min_{U_i\in\mathcal{U}_i}J_i\big(U_i,U_{-i}^{*}\big)}{\big|J_i\big(U_i^{*},U_{-i}^{*}\big)\big|}\;\ge 0 ,
\end{equation}
which is zero when the player cannot gain by changing its own plan.

\paragraph*{Statistical comparison}
Each comparison uses the per-agent difference in error, Nash minus the other model, so negative values favor Nash. Confidence intervals are obtained by bootstrapping \cite{cameron2015practitioner}. The $p$-values come from a sign-flip permutation test, which randomly reverses the signs of the differences to find how often a mean difference this large arises by chance. Because many comparisons are made, the $p$-values are Holm-adjusted \cite{holm1979simple} within three families: the position errors against SMIRL and SUMO (48 comparisons fixed before the analysis, including model variants not reported here), the four contrasts between the variants of our model, and the speed, acceleration, and jerk errors against SMIRL and SUMO (12 comparisons).

\paragraph*{Cross-validation}
To use all data, the comparison is repeated in a 10-fold cross-validation over the 370 training cases with the end location unknown. Each fold is predicted with group means from the other folds (SMIRL's reward is relearned per fold).

\section{Results}
\label{sec:results}

\subsection{Calibration}
\label{sec:calibration_results}
\begin{figure}[!t]
\centering
\includegraphics[width=0.88\columnwidth]{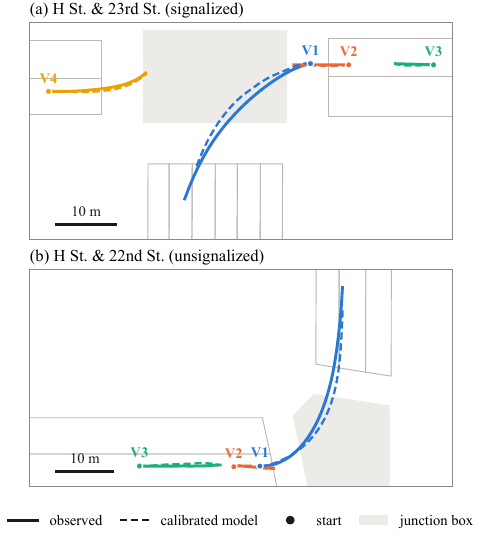}
\caption{Observed (solid) and calibrated (dashed) trajectories in two training cases; colors distinguish vehicles. (a) Signalized intersection, four vehicles, mean RMS error 0.81\,m. (b) Unsignalized intersection, three vehicles, mean RMS error 0.89\,m.}
\label{fig:nash-example}
\end{figure}

The calibrated game reproduces the trajectories of the 819 training agents with a mean time-aligned RMS error of 1.00\,m (median 0.89\,m): 1.00\,m at the signalized intersections and 0.94\,m at the unsignalized one. SUMO, calibrated with the same procedure, reaches 1.46\,m. Fig.~\ref{fig:nash-example} shows two typical cases.

The calibrated parameters are heterogeneous and reflect the maneuvers. At the signals, left-turning vehicles have the lowest desired speed (median 6.1\,m/s, against 7.3\,m/s for through vehicles) and the strongest speed tracking,; right-turning vehicles have the highest desired speed (8.6\,m/s) and steering weight; and through vehicles, which usually follow a leader, have the strongest gap keeping (median $w_g$ of 88). At the unsignalized intersection, the interaction weights are higher: through vehicles weigh conflict avoidance about three times more than at the signals (median $w_c$ of 199 versus 68), and gap keeping is stronger in every movement. With 3 to 22 interacting agents per unsignalized group, these differences are descriptive. The median perception delay of the 521 interacting agents is 1.31\,s (interquartile range 0.59--2.26\,s; group medians 1.27--1.41\,s at the signals and 1.60--2.32\,s without).

\paragraph*{Significance of the interaction terms and the perception delay}
For the 521 agents with a partner, the interaction terms reduce the error from 1.03 to 0.88\,m ($t=6.11$, $p<0.001$) and the perception delay from 0.93 to 0.88\,m ($t=3.46$, $p=0.001$), and both effects remain significant for all agents ($p<0.001$). On held-out agents with group-mean parameters, however, neither is significant: the delay is a driver-specific trait that a group average does not carry, which is why the held-out predictions use $\tau=0$.

\paragraph*{Practical identifiability of the calibrated parameters}
\label{sec:identifiability}
We assess which parameters the trajectories determine by local practical identifiability \cite{brun2001practical} on 60 training cases (120 agents): each parameter is set to nine values spanning its search interval, the others are kept at their calibrated values, and the game is re-solved. With the end point given (location and arrival time, as in calibration), a parameter is \emph{not identifiable} for an agent if the error does not rise by $\max(0.05\,\text{m}, 5\%)$ on either side of the calibrated value; with it not given, as in prediction, we report the error range and the share of agents with a \emph{clear minimum} inside the interval. Interaction parameters, evaluated only where their cost term is present, are also scored by the error near the closest approach, the closest distance, and the post-encroachment time (PET) \cite{allen1978analysis,gettman2003surrogate} (Table~\ref{tab:identifiability}).

\begin{table*}[!t]
\caption{Practical Identifiability of the Calibrated Parameters (120 Training Agents in 60 Cases)}
\label{tab:identifiability}
\centering
\footnotesize
\setlength{\tabcolsep}{4pt}
\resizebox{\textwidth}{!}{%
\begin{tabular}{@{}lr cc cc cc l@{}}
\hline
 & & \multicolumn{2}{c}{End point given} & \multicolumn{2}{c}{End point not given} & \multicolumn{2}{c}{Interaction measures} & \\
Parameter & Agents & Not identifiable & Error range (m) & Error range (m) & Clear minimum & Respond & Clear minimum & Assessment (reason)\\
\hline
$w_v$ & 120 & 0\% & 1.92 & 0.96 & 21\% & -- & -- & identifiable\\
$\bar v$ & 120 & 37\% & 0.29 & 6.65 & 32\% & -- & -- & weak in calibration (the end point fixes the arrival); dominant in prediction\\
$w_\kappa$ & 120 & 50\% & 0.15 & 0.12 & 2\% & -- & -- & weak (the lane cost fixes the path); matters in simulation\\
$w_a$ & 120 & 6\% & 0.88 & 0.94 & 10\% & -- & -- & identifiable\\
$b_p$ & 120 & 5\% & 1.49 & 1.77 & 12\% & -- & -- & identifiable\\
$b_v$ & 120 & 23\% & 0.28 & 0.40 & 2\% & -- & -- & weak (overlaps the speed-tracking cost)\\
$b_\theta$ & 120 & 96\% & 0.00 & 0.00 & 1\% & -- & -- & not identifiable (the lane cost fixes the heading)\\
$w_g$ & 28 & 21\% & 0.94 & 1.61 & 18\% & 79\% & 39\% & identifiable (requires a same-stream leader)\\
$w_c$ & 55 & 71\% & 0.00 & 0.06 & 0\% & 53\% & 11\% & not identifiable (term rarely active in the data); matters in simulation\\
$\tau$ & 68 & 50\% & 0.08 & 0.28 & 13\% & 68\% & 26\% & partly (acts only through the interaction terms)\\
\hline
\end{tabular}%
}
\vspace{2pt}

\parbox{0.98\textwidth}{\footnotesize Nine values per search interval, delayed-perception model, 50 iterations. Error range: median over agents. Clear minimum: interior minimum with a rise above $\max(0.05\,\text{m}, 5\%)$ on both sides. Interaction measures (end point not given): error within $\pm2$\,s of the closest approach, closest distance, and PET; an agent counts if any of them responds or has a clear minimum. $w_g$: agents with a leader; $w_c$: with a conflict partner; $\tau$: with either.}
\end{table*}

The trajectories identify the speed-tracking, acceleration, and terminal-position weights ($w_v$, $w_a$, $b_p$; error range 0.9--1.9\,m, identifiable for at least 94\% of agents) and the gap-keeping weight $w_g$ of followers, which changes the PET by a median of 0.95\,s. The desired speed $\bar v$ and the delay $\tau$ are identified only in prediction: $\bar v$ matters little when the end point fixes the arrival but becomes the dominant parameter once it is free (median range 6.65\,m), and $\tau$ changes an interaction measure for 68\% of the interacting agents. The remaining parameters are not identified by position: $b_v$ and $b_\theta$ overlap with the speed and lane costs, $w_\kappa$ changes how smoothly a vehicle turns, not where, and $w_c$ is rarely activated because recorded crossings stay far apart (median predicted separation 9--13\,m, against $R_c=3$\,m). Yet both govern the simulated behavior once activated, as shown next.

\paragraph*{Weakly identified parameters in closed-loop simulation}
\label{sec:sensitivity}
To examine what $w_c$ and $w_\kappa$ do in a simulator, we build 93 negotiation scenarios from 12 recorded vehicle pairs on crossing or merging routes, shifting one vehicle so that the nominal arrivals at the conflict point differ by $-2$ to $+2$\,s, far tighter than in the recordings. Both vehicles are controlled by the game in closed loop, with 5\,s plans re-solved every 0.5\,s in receding horizon \cite{wang2021game}, and all parameters at their group means except the varied one, set equal for every vehicle (819 runs). Table~\ref{tab:sensitivity} reports the results.

\begin{table}[!t]
\caption{Closed-Loop Effect of $w_c$, $w_\kappa$, and the Contact Barrier in 93 Negotiation Scenarios}
\label{tab:sensitivity}
\centering
\footnotesize
\setlength{\tabcolsep}{3pt}
\resizebox{\columnwidth}{!}{%
\begin{tabular}{@{}lccccc@{}}
\hline
Value & Collisions & Min.\ gap (m) & PET (s) & Peak decel.\ (m/s$^2$) & Peak lat.\ acc.\ (m/s$^2$)\\
\hline
\multicolumn{6}{@{}l}{\emph{Conflict weight $w_c$}}\\
10 & 48\% & $-0.15$ & 1.0 & 2.54 & 3.04\\
31.6 & 47\% & $-0.01$ & 1.0 & 2.60 & 3.05\\
100 & 43\% & $0.39$ & 1.1 & 2.51 & 3.03\\
316 & 28\% & $0.93$ & 1.2 & 2.29 & 3.06\\
1000 & 12\% & $2.11$ & 1.3 & 1.86 & 3.13\\
\hline
\multicolumn{6}{@{}l}{\emph{Curvature weight $w_\kappa$}}\\
0.1 & 32\% & $0.89$ & 1.15 & 2.22 & 4.05\\
1 & 34\% & $0.81$ & 1.10 & 2.43 & 4.02\\
10 & 35\% & $0.72$ & 1.10 & 2.39 & 3.08\\
100 & 28\% & $1.06$ & 1.10 & 1.96 & 1.89\\
\hline
\multicolumn{6}{@{}l}{\emph{Both set: $w_c=1000$, $w_\kappa=100$}}\\
 & 12\% & $2.36$ & 1.40 & 1.54 & 1.89\\
\hline
\multicolumn{6}{@{}l}{\emph{Both set, rectangular barrier \eqref{eq:contact-box}, $W_c=10^5$}}\\
 & \textbf{0\%} & $2.18$ & 1.40 & 1.69 & 1.94\\
\hline
\end{tabular}%
}
\vspace{2pt}

\parbox{\columnwidth}{\footnotesize Nash with zero delay, receding horizon, end point not given, group-mean parameters except the varied one. Medians over scenarios, except collisions (share of scenarios with overlapping outlines). Min.\ gap: bumper-to-bumper; negative means overlap. Peak deceleration: larger of the two vehicles; lateral acceleration: turning vehicles. All rows but the last use the elliptical barrier \eqref{eq:contact} ($W_c=1000$). No deadlocks.}
\end{table}

\emph{Conflict weight.} The conflict term keeps vehicles apart in tight conflicts. Raising $w_c$ from 100 to 1000 cuts collisions from 43\% to 12\% and widens the gaps while softening the braking, because the vehicles adjust earlier; collisions fall in 10 of the 12 pairs and vanish in 6. The remaining collisions occur because the elliptical barrier with $W_c=1000$ is bounded: near contact, accepting its cost is cheaper than braking.

\emph{Curvature weight.} Over its interval, $w_\kappa$ lowers the peak lateral acceleration in turns from 4.05 to 1.89\,m/s$^2$, in every pair. The observed turns (490 vehicles) have a median peak of 1.59\,m/s$^2$ (interquartile range 1.15--2.04), which matches $w_\kappa=100$.

\emph{Setting the weakly identified weights.} These results suggest estimating from the trajectories the parameters they identify and setting the others from safety and comfort targets. Setting $w_c=1000$ and $w_\kappa=100$ for every vehicle, with all other parameters at their group means (second-to-last row of Table~\ref{tab:sensitivity}), reduces collisions from 43\% to 12\%, widens the gaps, softens the braking, and makes turns human-like (1.89\,m/s$^2$). Accuracy improves at the same time, by 0.19\,m on the held-out cases with the end point given (95\% CI $[-0.33,-0.09]$) and from 7.45 to 7.29\,m in the cross-validation over all 819 agents ($p<0.001$).

\emph{A collision-free option.} Replacing the elliptical barrier with the rectangular barrier \eqref{eq:contact-box} with $W_c=10^5$ removes every collision (0 of 93 scenarios; last row of Table~\ref{tab:sensitivity}) without deadlock or refitting. The interaction remains human-like (90th-percentile peak deceleration 2.65\,m/s$^2$); in the hardest scenario a vehicle brakes at the 6\,m/s$^2$ bound instead of colliding. Acting only near contact, the barrier is a safety layer on top of the calibrated driving style. On the held-out cases it leaves the accuracy of the weights set from safety and comfort unchanged ($-0.01$\,m with the end point given, 95\% CI $[-0.02,+0.00]$; $-0.10$\,m without it, $[-0.28,+0.05]$), although more solves stop at the 50-iteration limit before converging.

\subsection{Held-out comparison of the five models}
\label{sec:res_benchmark}
\begin{figure*}[!t]
\centering
\includegraphics[width=\textwidth]{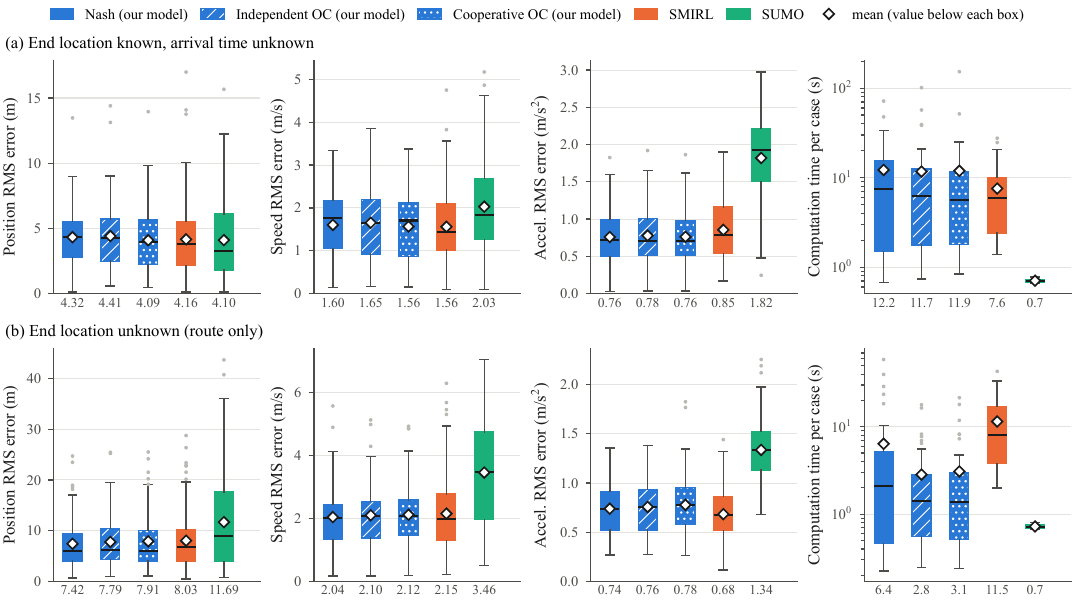}
\caption{Held-out comparison (42 cases, 96 agents): per-agent RMS errors of time-aligned position, speed, and acceleration, and single-thread computation time per case (right, logarithmic). Nash, independent OC, and cooperative OC are three couplings of our model (blue). Boxes show the quartiles, whiskers extend to 1.5 times the interquartile range, and diamonds mark the means, whose values are printed below the boxes. (a) End location known, arrival time unknown. (b) End location unknown.}
\label{fig:comparison}
\end{figure*}

Fig.~\ref{fig:comparison} and Table~\ref{tab:benchmark} report the errors on the 96 held-out agents; paired differences are Nash minus the other model, so negative values favor the game.

\begin{figure*}[!t]
\centering
\includegraphics[width=\textwidth]{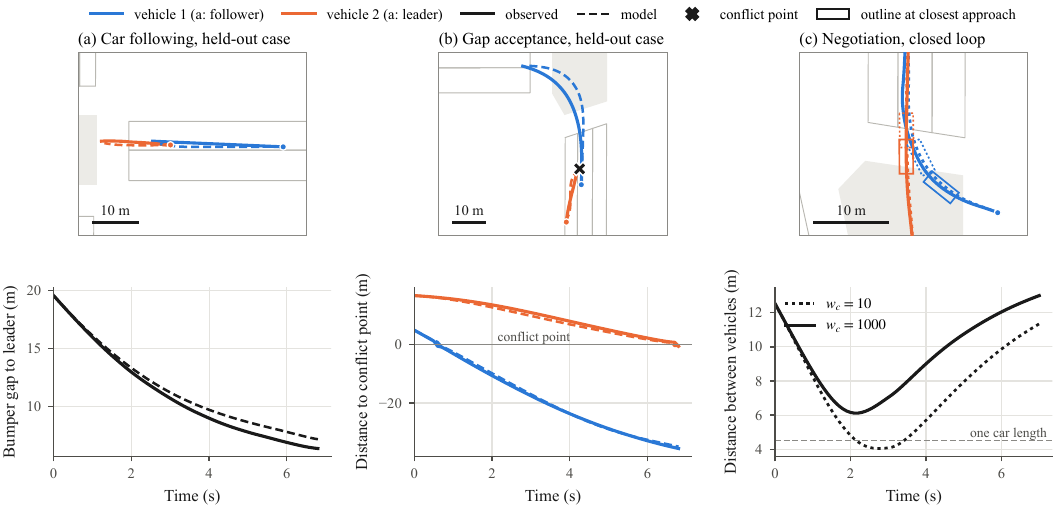}
\caption{Interaction behaviors captured by the model. Top: paths on the intersection map; bottom: the interaction over time. (a) Car following in a held-out case: the predicted bumper gap to the leader follows the observed one as the follower closes up (RMS gap error 0.63\,m). (b) Gap acceptance in a held-out case: the second vehicle approaches slowly and enters the conflict point after the first has cleared it; the observed post-encroachment time of 6.2\,s is predicted as 6.1\,s. (c) Negotiation in a closed-loop scenario (merging, simultaneous nominal arrival): with a weak conflict term ($w_c=10$, dotted) the vehicles come closer than one car length and their outlines overlap; with $w_c=1000$ (solid) the merging vehicle adjusts early and the minimum distance is 6\,m. The outlines mark both vehicles at their closest approach.}
\label{fig:heldout-cases}
\end{figure*}

\begin{table}[!t]
\caption{Held-Out Mean RMS Errors of the Five Models (96 Agents) and Close Approaches (90 Agent Pairs)}
\label{tab:benchmark}
\centering
\footnotesize
\setlength{\tabcolsep}{3pt}
\resizebox{\columnwidth}{!}{%
\begin{tabular}{@{}lccccccc@{}}
\hline
 & & & & & \multicolumn{2}{c}{Position} & Close\\
Model & Position & Speed & Accel. & Jerk & partner & no partner & appr.\\
\hline
\multicolumn{8}{@{}l}{\emph{(1) End location known, arrival time unknown}}\\
Nash (our model) & 4.325 & 1.602 & \textbf{0.759} & 0.868 & 4.471 & 4.102 & 1\\
Independent (our model) & 4.414 & 1.653 & 0.778 & 0.871 & 4.605 & 4.121 & 1\\
Cooperative (our model) & \textbf{4.089} & 1.565 & 0.761 & \textbf{0.866} & \textbf{4.193} & 3.931 & 0\\
SMIRL & 4.161 & \textbf{1.559} & 0.855 & 1.304 & 4.536 & 3.587 & 4\\
SUMO & 4.104 & 2.028 & 1.820 & 6.780 & 4.586 & \textbf{3.367} & 0\\
\hline
\multicolumn{8}{@{}l}{\emph{(2) End location unknown}}\\
Nash (our model) & \textbf{7.420} & \textbf{2.043} & 0.739 & 0.866 & \textbf{8.110} & \textbf{6.367} & 0\\
Independent (our model) & 7.786 & 2.100 & 0.757 & 0.873 & 8.680 & 6.422 & 0\\
Cooperative (our model) & 7.907 & 2.118 & 0.778 & 0.899 & 8.868 & 6.440 & 0\\
SMIRL & 8.032 & 2.153 & \textbf{0.683} & \textbf{0.779} & 8.360 & 7.532 & 5\\
SUMO & 11.689 & 3.456 & 1.336 & 4.279 & 12.956 & 9.755 & 2.2\\
\hline
\end{tabular}%
}
\vspace{2pt}

\parbox{\columnwidth}{\footnotesize Position in m, speed in m/s, acceleration in m/s$^2$, jerk in m/s$^3$. Partner: the 58 agents with an interaction partner within 20\,m. Close approaches: agent pairs whose centers come within 2\,m (SUMO: mean over its 10 seeds); none occurs in the recording.}
\end{table}

\paragraph*{End location known, arrival time unknown}
A given end location fixes most of the motion, so position accuracy no longer separates the models: all five reach mean errors of 4.09--4.41\,m, and no difference is significant. SUMO has the lowest median position error (3.28\,m, against 4.31\,m for Nash), but its mean differs from that of the game by only 0.22\,m (95\% CI $[-0.58,1.00]$, Holm $p=1.00$). How a vehicle reaches its end point, however, shows in the speed and acceleration errors (Fig.~\ref{fig:comparison}). The game follows the observed speed 21\% more closely than SUMO (Holm $p=0.003$) and the acceleration 58\% more closely (Holm $p<0.001$; lower for 93 of the 96 agents), and SUMO's jerk error is nearly eight times that of the game (higher for every agent), a result of its piecewise-constant accelerations. The game also has lower acceleration and jerk errors than SMIRL (by 11\% and 33\%, Holm $p=0.037$ and $p<0.001$), with equal speed errors.

\paragraph*{End location unknown}
When only the route is known, the game has the lowest position and speed errors of all five models (7.42\,m, 2.04\,m/s). It is 4.27\,m more accurate than SUMO (Holm $p=0.002$), whose speed and acceleration errors it reduces by more than 40\% (Holm $p<0.001$), and more accurate than independent OC by 0.37\,m (Holm $p=0.031$), cooperative OC by 0.49\,m (Holm $p=0.014$), and SMIRL by 0.61\,m (95\% CI $[-1.18,-0.01]$).

\paragraph*{Cross-validation}
When all 819 training agents are predicted out of fold (Table~\ref{tab:crossval}), the ranking is the same: the game is more accurate than independent OC by 0.33\,m, cooperative OC by 0.32\,m, SMIRL by 0.93\,m, and SUMO by 3.61\,m (all $p<0.001$).

\begin{table}[!t]
\caption{10-Fold Cross-Validation Over the 370 Calibration Cases, End Location Unknown: Position RMS Error (m) and Paired Difference, Nash Minus the Other Model}
\label{tab:crossval}
\centering
\footnotesize
\setlength{\tabcolsep}{3pt}
\resizebox{\columnwidth}{!}{%
\begin{tabular}{@{}lccc@{}}
\hline
Model & All (819) & With partner (521) & Without partner (298)\\
\hline
Nash (our model) & \textbf{7.45} & \textbf{7.62} & \textbf{7.16}\\
Independent (our model) & 7.78 & 8.09 & 7.23\\
Cooperative (our model) & 7.77 & 8.04 & 7.31\\
SMIRL & 8.38 & 8.55 & 8.10\\
SUMO & 11.06 & 10.68 & 11.72\\
\hline
Nash $-$ Independent & $-0.33$ [$-0.44$, $-0.22$] & $-0.47$ [$-0.65$, $-0.31$] & $-0.07$ [$-0.14$, $-0.00$]\\
Nash $-$ Cooperative & $-0.32$ [$-0.45$, $-0.20$] & $-0.42$ [$-0.63$, $-0.24$] & $-0.15$ [$-0.23$, $-0.07$]\\
Nash $-$ SMIRL & $-0.93$ [$-1.25$, $-0.63$] & $-0.93$ [$-1.35$, $-0.53$] & $-0.94$ [$-1.24$, $-0.63$]\\
Nash $-$ SUMO & $-3.61$ [$-4.37$, $-2.90$] & $-3.07$ [$-4.02$, $-2.21$] & $-4.55$ [$-5.57$, $-3.59$]\\
\hline
\end{tabular}%
}
\vspace{2pt}

\parbox{\columnwidth}{\footnotesize Group means (for SMIRL, a reward) estimated without the case's fold. Nash with zero delay, 50 iterations. Partner: leader or conflict partner within 20\,m. 95\% CIs by bootstrap; all differences $p<0.001$ (sign-flip test) except Nash $-$ Independent without partner ($p=0.041$).}
\end{table}

\paragraph*{Effect of strategic coupling}
The advantage of the game appears where the coupling acts: in the cross-validation, it amounts to 0.47\,m over independent OC and 0.42\,m over cooperative OC for the 521 agents with an interaction partner, but only 0.07 and 0.15\,m for the 298 agents without one (Table~\ref{tab:crossval}). The drivers in these data are therefore best described as anticipating each other while pursuing their own objectives. The game also reproduces the spacing within vehicle pairs better than cooperative OC (7.26 against 7.71\,m; 95\% CI $[-0.75,-0.11]$\,m). All three models fit the training trajectories almost equally well (1.00, 1.03, and 1.05\,m), so the choice of solution concept becomes visible in prediction rather than in fit.

\paragraph*{Effect of driver heterogeneity}
Each held-out vehicle receives either the \emph{group mean}, the mean parameter vector of the calibrated training drivers of its group (signal control $\times$ movement), or a \emph{correlated draw}, the complete parameter vector of one calibrated training driver of the group, which preserves the correlations between the parameters. The group mean is the best estimate for a vehicle whose own parameters are unknown: it predicts speed and acceleration more accurately in both settings, and position when the end point and arrival time are given (2.47 against 2.68\,m; $p=0.13$; Table~\ref{tab:heterogeneity}a). A draw adds the difference between the drawn driver and the observed one, yet with the end point unknown it is 0.39\,m more accurate in position, although not significantly ($p=0.25$). Because the trajectory depends nonlinearly on the parameters, the trajectory computed with the averaged parameters differs from the average of the trajectories of the individual drivers: with the end point unknown, vehicles given the group mean travel at 3.72\,m/s on average, faster than the recorded 3.20\,m/s, whereas vehicles given correlated draws travel at 3.00\,m/s (Table~\ref{tab:heterogeneity}b). The data show heterogeneity at two levels. Between groups, the group means are 0.13\,m more accurate than one pooled vector for all drivers when the end point is unknown ($p=0.004$). Within groups, with the end point unknown, identical drivers reproduce only 81\% of the recorded spread of mean speeds across vehicles, whereas correlated draws reproduce 92\%; with the end point and time given, the recorded arrival time largely fixes each vehicle's mean speed, so both assignments match the recorded speeds. A simulator should therefore use the group mean to predict a particular vehicle and correlated draws to generate a realistic population of drivers.

\begin{table}[!t]
\caption{Held-Out Errors and Speeds of the Game Under Different Parameter Assignments (96 Agents)}
\label{tab:heterogeneity}
\centering
\footnotesize
\setlength{\tabcolsep}{3pt}
\resizebox{\columnwidth}{!}{%
\begin{tabular}{@{}lcccccc@{}}
\hline
 & \multicolumn{3}{c}{End point and time given} & \multicolumn{3}{c}{End point unknown}\\
\multicolumn{7}{@{}l}{\emph{(a) RMS error per vehicle}}\\
Parameters & Pos. & Speed & Accel. & Pos. & Speed & Accel.\\
\hline
Group mean & \textbf{2.47} & \textbf{1.02} & \textbf{0.68} & 7.43 & \textbf{2.04} & \textbf{0.74}\\
Correlated draw & 2.68 & 1.14 & 0.79 & \textbf{7.04} & 2.15 & 0.89\\
\hline
 & Mean & SD & Ratio & Mean & SD & Ratio\\
\hline
Recorded & 3.20 & 1.67 & 1 & 3.20 & 1.67 & 1\\
Group mean & 3.21 & 1.67 & 1.00 & 3.72 & 1.36 & 0.81\\
Correlated draw & 2.99 & 1.63 & 0.97 & 3.00 & 1.54 & 0.92\\
\hline
\end{tabular}%
}
\vspace{2pt}

\parbox{\columnwidth}{\footnotesize (a) Position (m), speed (m/s), acceleration (m/s$^2$). (b) Each vehicle's mean speed over its window (m/s): average over the vehicles, standard deviation (SD) across them, and SD relative to the recorded SD. Ten draws per vehicle, averaged; zero delay, 50 iterations.}
\end{table}

\paragraph*{Interpreting the size of the errors}
The 7.42\,m error of the game in Table~\ref{tab:benchmark} is that of a single prediction from the initial state; its ADE and FDE (0.91 and 2.45\,m at 3\,s, 3.02 and 8.41\,m at 6\,s) are comparable to single-prediction forecasting models, which see an observed history but not the route (ADE(6\,s) 3.69--4.61\,m and FDE(6\,s) 8.65--11.21\,m on nuScenes \cite{messaoud2021trajectory}; ADE(3\,s) 1.71\,m on Argoverse \cite{liang2020learning}). About 90\% of the squared error lies along the path, so the error is mainly one of timing, part of which stems from the unrecorded signal phases.

\paragraph*{Interaction behaviors captured by the model}
Fig.~\ref{fig:heldout-cases} illustrates three interaction behaviors: car following and gap acceptance in two held-out cases with transferred parameters, and negotiation in a closed-loop scenario. With the end point given, the model predicts the bumper gap of the 35 car-following pairs in the held-out set with a median RMS error of 2.1\,m, and it reproduces the passing order of 7 of the 8 gap-acceptance pairs with a median PET error of 0.25\,s.

\paragraph*{Safety}
Among the 90 held-out agent pairs, the game produced at most one close approach, against four and five for SMIRL, which plans against constant-velocity predictions of the others (Table~\ref{tab:benchmark}).

\subsection{Computation time and convergence}
\label{sec:res_time}
\begin{table}[!t]
\caption{Single-Thread Computation Time per Held-Out Case (42 Cases) and Convergence of the Optimization-Based Models (96 Agents)}
\label{tab:timing}
\centering
\footnotesize
\setlength{\tabcolsep}{3pt}
\resizebox{\columnwidth}{!}{%
\begin{tabular}{@{}lccccccc@{}}
\hline
 & \multicolumn{4}{c}{Computation time} & \multicolumn{3}{c}{Agents by solver outcome}\\
Model & Mean (s) & Median (s) & 90th pct. (s) & RTF & Conv. & Budget & Stalled\\
\hline
\multicolumn{8}{@{}l}{\emph{(1) End location known, arrival time unknown}}\\
Nash (our model) & 12.20 & 7.42 & 30.83 & 0.49 & 68 & 15 & 13\\
Independent (our model) & 11.69 & 6.25 & 20.75 & 0.42 & 63 & 17 & 16\\
Cooperative (our model) & 11.92 & 5.61 & 20.05 & 0.37 & 76 & 14 & 6\\
SMIRL & 7.56 & 5.97 & 14.86 & 0.71 & -- & -- & --\\
SUMO & 0.71 & 0.71 & 0.74 & 0.09 & -- & -- & --\\
\hline
\multicolumn{8}{@{}l}{\emph{(2) End location unknown}}\\
Nash (our model) & 6.38 & 2.08 & 17.57 & 0.25 & 79 & 6 & 11\\
Independent (our model) & 2.84 & 1.40 & 6.90 & 0.16 & 88 & 3 & 5\\
Cooperative (our model) & 3.08 & 1.38 & 8.12 & 0.16 & 88 & 3 & 5\\
SMIRL & 11.48 & 8.12 & 22.86 & 1.00 & -- & -- & --\\
SUMO & 0.72 & 0.72 & 0.78 & 0.09 & -- & -- & --\\
\hline
\end{tabular}%
}
\vspace{2pt}

\parbox{\columnwidth}{\footnotesize RTF: median over cases of computation time divided by simulated horizon. SUMO: one run (network construction adds 0.15\,s). Conv.: tolerances of Section~\ref{sec:numerical_stability} met; Budget: 100 iterations reached while still decreasing; Stalled: no descent step, best iterate kept.}
\end{table}

With the end location unknown (Table~\ref{tab:timing}), the game needs a median of 2.1\,s per case (real-time factor 0.25), a quarter of the time of SMIRL (8.1\,s); SUMO needs 0.7\,s. The time grows from 0.4\,s for one vehicle to 1.9, 5.2, and 10.4\,s for two, three, and four, so typical interactions are solved faster than real time even with unoptimized code. With the end location known, the 15\,s horizon raises the median to 7.4\,s (real-time factor 0.49). The three optimization-based models converge at similar rates, so the coupling does not make convergence much harder.

\paragraph*{Equilibrium verification}
\label{sec:res_gap}
For converged held-out solves, the median relative Nash gap \eqref{eq:nashgap} is below $10^{-6}$ (90th percentile below 0.6\%), and 93--94\% of the players cannot lower their own cost by more than 1\% through a unilateral change (86--89\% including unconverged solves), so the solutions are approximate local equilibria. Plans of independent OC, evaluated in the game, have gaps 50 to 250 times larger, with only 58--60\% of players within 1\%. On the cases in which every player's gap is below 1\%, the results are unchanged ($-0.38$, $-0.48$, and $-4.04$\,m against independent OC, cooperative OC, and SUMO, end location unknown).

\section{Conclusion}
\label{sec:conclusion}
This study calibrated a differential game of vehicle interactions at urban intersections in which every driver pursues its own objectives, anticipates the others, and perceives them with a delay. Calibrated on 819 vehicles, it reproduces their trajectories with a mean error of 1.00\,m. When only the route is known, it predicts held-out interactions more accurately than calibrated SUMO, IRL, and the independent and cooperative variants, with cross-validated error reductions of 3.61, 0.93, 0.33, and 0.32\,m (all $p<0.001$). The advantage of the Nash coupling is concentrated among interacting vehicles, so mutual anticipation among self-interested drivers matters for prediction.

For game-theoretic simulators, the identifiability analysis suggests estimating from trajectories only the parameters they identify and setting the conflict and steering weights from safety and comfort targets; with a stiffer contact barrier, these settings removed every collision in tight closed-loop conflicts. Drawing each simulated driver's parameters from the calibrated drivers of its group reproduces the variability of observed traffic.

Future work will embed the game in a full simulator in which interaction groups form and dissolve, validate it at further sites with signal-phase data, calibrate it on interaction-rich data with closer conflicts so that the conflict weight and the reaction delay become identifiable, and extend it to pedestrians, cyclists, and automated vehicles.

\section*{Author Contributions}

A. Khakpour: Conceptualization, Methodology, Software, Data curation, Model calibration, Validation, Visualization, Writing -- original draft, Writing -- review \& editing.
N. Li: Conceptualization, Writing -- review \& editing.
P. Beigi: Data curation, Model calibration, Writing -- review \& editing.
Y. (M.) Nie: Supervision, Writing -- review \& editing.

\section*{Acknowledgment}

This research was conducted with support from the Center for Connected and Automated Transportation (CCAT), a regional University Transportation Center sponsored by the U.S. Department of Transportation (USDOT). The authors thank Hussein Kobeissi for help in porting part of the iterative linear--quadratic game code.
The authors honor the memory of their mentor and colleague, Professor Hani S. Mahmassani, who passed away after contributing to the early stages of this study. His ideas and thoughtful guidance were instrumental in shaping the research direction and strengthening its conceptual foundation. His scholarship, leadership, and generosity as a mentor have left a lasting influence on the authors and the transportation research community. This paper is dedicated to his memory and enduring legacy.

\bibliographystyle{IEEEtran}
\bibliography{references}

@article{chandler_traffic_1958,
  author  = {Chandler, Robert E. and Herman, Robert and Montroll, Elliott W.},
  title   = {Traffic Dynamics: Studies in Car Following},
  journal = {Operations Research},
  volume  = {6},
  number  = {2},
  pages   = {165--184},
  year    = {1958},
  doi     = {10.1287/opre.6.2.165}
}

@article{gazis_car-following_1959,
  author  = {Gazis, Denos C. and Herman, Robert and Potts, Renfrey B.},
  title   = {Car-Following Theory of Steady-State Traffic Flow},
  journal = {Operations Research},
  volume  = {7},
  number  = {4},
  pages   = {499--505},
  year    = {1959},
  doi     = {10.1287/opre.7.4.499}
}

@article{kesting_enhanced_2010,
  author  = {Kesting, Arne and Treiber, Martin and Helbing, Dirk},
  title   = {Enhanced Intelligent Driver Model to Assess the Impact of Driving Strategies on Traffic Capacity},
  journal = {Philosophical Transactions of the Royal Society A: Mathematical, Physical and Engineering Sciences},
  volume  = {368},
  number  = {1928},
  pages   = {4585--4605},
  year    = {2010},
  doi     = {10.1098/rsta.2010.0084}
}

@article{kesting2007general,
  author  = {Kesting, Arne and Treiber, Martin and Helbing, Dirk},
  title   = {General Lane-Changing Model {MOBIL} for Car-Following Models},
  journal = {Transportation Research Record},
  volume  = {1999},
  number  = {1},
  pages   = {86--94},
  year    = {2007},
  doi     = {10.3141/1999-10}
}

@article{gettman2003surrogate,
  author  = {Gettman, Douglas and Head, Larry},
  title   = {Surrogate Safety Measures from Traffic Simulation Models},
  journal = {Transportation Research Record},
  volume  = {1840},
  number  = {1},
  pages   = {104--115},
  year    = {2003},
  doi     = {10.3141/1840-12}
}

@article{barmpounakis2016modeling,
  author  = {Barmpounakis, Emmanouil N. and Vlahogianni, Eleni I. and Golias, John C.},
  title   = {Modeling Cooperation and Powered-Two-Wheeler Short-Term Strategic Decisions During Overtaking in Urban Arterials},
  journal = {International Journal of Transportation Science and Technology},
  volume  = {5},
  number  = {4},
  pages   = {227--238},
  year    = {2016}
}

@article{kita1999merging,
  author  = {Kita, Hideyuki},
  title   = {A Merging--Giveway Interaction Model of Cars in a Merging Section: A Game-Theoretic Analysis},
  journal = {Transportation Research Part A: Policy and Practice},
  volume  = {33},
  number  = {3--4},
  pages   = {305--312},
  year    = {1999},
  doi     = {10.1016/S0965-8564(98)00039-1}
}

@article{talebpour2015modeling,
  author  = {Talebpour, Alireza and Mahmassani, Hani S. and Hamdar, Samer H.},
  title   = {Modeling Lane-Changing Behavior in a Connected Environment: A Game Theory Approach},
  journal = {Transportation Research Procedia},
  volume  = {7},
  pages   = {420--440},
  year    = {2015},
  doi     = {10.1016/j.trpro.2015.06.022}
}

@book{isaacs1965differential,
  author    = {Isaacs, Rufus},
  title     = {Differential Games: A Mathematical Theory with Applications to Warfare and Pursuit, Control and Optimization},
  publisher = {John Wiley \& Sons},
  address   = {New York, NY},
  year      = {1965}
}

@book{basar1999dynamic,
  author    = {Ba{\c{s}}ar, Tamer and Olsder, Geert Jan},
  title     = {Dynamic Noncooperative Game Theory},
  edition   = {2},
  publisher = {Society for Industrial and Applied Mathematics},
  address   = {Philadelphia, PA},
  year      = {1999},
  doi       = {10.1137/1.9781611971132}
}

@article{bressan2011noncooperative,
  author  = {Bressan, Alberto},
  title   = {Noncooperative Differential Games},
  journal = {Milan Journal of Mathematics},
  volume  = {79},
  number  = {2},
  pages   = {357--427},
  year    = {2011}
}

@article{li2020game,
  author  = {Li, Nan and Yao, Yiqi and Kolmanovsky, Ilya and Atkins, Ella and Girard, Anouck and Yildiz, Yildiray},
  title   = {Game-Theoretic Modeling of Multi-Vehicle Interactions at Uncontrolled Intersections},
  journal = {IEEE Transactions on Intelligent Transportation Systems},
  volume  = {23},
  number  = {2},
  pages   = {1428--1442},
  year    = {2022},
  doi     = {10.1109/TITS.2020.3026160}
}

@article{hang2022driving,
  author  = {Hang, Peng and Huang, Chao and Hu, Zhongxu and Lv, Chen},
  title   = {Driving Conflict Resolution of Autonomous Vehicles at Unsignalized Intersections: A Differential Game Approach},
  journal = {IEEE/ASME Transactions on Mechatronics},
  volume  = {27},
  number  = {6},
  pages   = {5136--5146},
  year    = {2022},
  doi     = {10.1109/TMECH.2022.3174273}
}

@article{fang2023cooperative,
  author  = {Fang, Shiyu and Hang, Peng and Wei, Chongfeng and Xing, Yang and Sun, Jian},
  title   = {Cooperative Driving of Connected Autonomous Vehicles in Heterogeneous Mixed Traffic: A Game Theoretic Approach},
  journal = {IEEE Transactions on Intelligent Vehicles},
  year    = {2024},
  doi     = {10.1109/TIV.2024.3399694}
}

@misc{davis2019merging,
  author        = {Davis, L. C.},
  title         = {Optimal Merging from an On-Ramp into a High-Speed Lane Dedicated to Connected Autonomous Vehicles},
  year          = {2018},
  eprint        = {1809.01226},
  archivePrefix = {arXiv},
  primaryClass  = {eess.SY}
}

@article{zhao_two-dimensional_2020,
  author  = {Zhao, Jing and Knoop, Victor L. and Wang, Meng},
  title   = {Two-Dimensional Vehicular Movement Modelling at Intersections Based on Optimal Control},
  journal = {Transportation Research Part B: Methodological},
  volume  = {138},
  pages   = {1--22},
  year    = {2020},
  doi     = {10.1016/j.trb.2020.04.001}
}

@inproceedings{sadigh2016planning,
  author    = {Sadigh, Dorsa and Sastry, Shankar and Seshia, Sanjit A. and Dragan, Anca D.},
  title     = {Planning for Autonomous Cars That Leverage Effects on Human Actions},
  booktitle = {Proceedings of Robotics: Science and Systems XII},
  pages     = {1--9},
  year      = {2016},
  doi       = {10.15607/RSS.2016.XII.016}
}

@article{fisac2019hierarchical,
  author  = {Fisac, Jaime F. and Bronstein, Eli and Uretsky, Erez and Sadigh, Dorsa and Sastry, Shankar and Russell, Stuart and Dragan, Anca D.},
  title   = {Hierarchical Game-Theoretic Planning for Autonomous Vehicles},
  journal = {IEEE Robotics and Automation Letters},
  volume  = {4},
  number  = {2},
  pages   = {1532--1539},
  year    = {2019},
  doi     = {10.1109/LRA.2019.2892654}
}

@article{li2019differential,
  author  = {Li, Na and Kolmanovsky, Ilya and Girard, Anouck},
  title   = {Differential Dynamic Games for Intelligent Vehicle Control},
  journal = {Annual Reviews in Control},
  volume  = {47},
  pages   = {51--66},
  year    = {2019},
  doi     = {10.1016/j.arcontrol.2019.04.001}
}

@article{schwarting2019social,
  author  = {Schwarting, Wilko and Pierson, Alyssa and Alonso-Mora, Javier and Karaman, Sertac and Rus, Daniela},
  title   = {Social Behavior for Autonomous Vehicles},
  journal = {Proceedings of the National Academy of Sciences},
  volume  = {116},
  number  = {24},
  pages   = {11669--11674},
  year    = {2019},
  doi     = {10.1073/pnas.1820676116}
}

@inproceedings{fridovich2020efficient,
  author    = {Fridovich-Keil, David and Ratner, Ellis and Peters, Lasse and Dragan, Anca D. and Tomlin, Claire J.},
  title     = {Efficient Iterative Linear-Quadratic Approximations for Nonlinear Multi-Player General-Sum Differential Games},
  booktitle = {Proceedings of the IEEE International Conference on Robotics and Automation},
  pages     = {1475--1481},
  publisher = {IEEE},
  year      = {2020},
  doi       = {10.1109/ICRA40945.2020.9197129}
}

@book{bryson1975applied,
  author    = {Bryson, Arthur E. and Ho, Yu-Chi},
  title     = {Applied Optimal Control: Optimization, Estimation, and Control},
  publisher = {Taylor \& Francis},
  year      = {1975}
}

@inproceedings{polack2017kinematic,
  author    = {Polack, Philip and Altch{\'e}, Florent and d'Andr{\'e}a-Novel, Brigitte and de La Fortelle, Arnaud},
  title     = {The Kinematic Bicycle Model: A Consistent Model for Planning Feasible Trajectories for Autonomous Vehicles?},
  booktitle = {Proceedings of the IEEE Intelligent Vehicles Symposium},
  pages     = {812--818},
  publisher = {IEEE},
  year      = {2017},
  doi       = {10.1109/IVS.2017.7995816}
}

@inproceedings{snoek2012practical,
  author    = {Snoek, Jasper and Larochelle, Hugo and Adams, Ryan P.},
  title     = {Practical Bayesian Optimization of Machine Learning Algorithms},
  booktitle = {Advances in Neural Information Processing Systems},
  volume    = {25},
  pages     = {2951--2959},
  year      = {2012}
}

@article{ammourah2025introduction,
  author  = {Ammourah, Rami and Beigi, Pedram and Fan, Bingyi and Hamdar, Samer H. and Hourdos, John and Hsiao, Chun-Chien and James, Rachel and Khajeh-Hosseini, Mohammdreza and Mahmassani, Hani S. and Monzer, Dana and Radvand, Tina and Talebpour, Alireza and Yousefi, Mahdi and Zhang, Yanlin},
  title   = {Introduction to the Third Generation Simulation Dataset: Data Collection and Trajectory Extraction},
  journal = {Transportation Research Record},
  volume  = {2679},
  number  = {1},
  pages   = {1768--1784},
  year    = {2025},
  doi     = {10.1177/03611981241257257}
}

@article{ni2020limitations,
  title={Limitations of Current Traffic Models and Strategies to Address Them},
  author={Ni, Daiheng},
  journal={Simulation Modelling Practice and Theory},
  volume={104},
  pages={102137},
  year={2020},
  doi={10.1016/j.simpat.2020.102137}
}

@article{zhao2022microscopic,
  title={Microscopic Traffic Modeling Inside Intersections: Interactions Between Drivers},
  author={Zhao, Jing and Knoop, Victor L. and Wang, Meng},
  journal={Transportation Science},
  volume={57},
  number={1},
  pages={135--155},
  year={2022},
  doi={10.1287/trsc.2022.1163}
}

@article{anvari2015modelling,
  title={Modelling Shared Space Users via Rule-Based Social Force Model},
  author={Anvari, Bani and Bell, Michael G. H. and Sivakumar, Aruna and Ochieng, Washington Y.},
  journal={Transportation Research Part C: Emerging Technologies},
  volume={51},
  pages={83--103},
  year={2015},
  doi={10.1016/j.trc.2014.10.012}
}

@article{treiber2000congested,
  title={Congested traffic states in empirical observations and microscopic simulations},
  author={Treiber, Martin and Hennecke, Ansgar and Helbing, Dirk},
  journal={Physical Review E},
  volume={62},
  number={2},
  pages={1805--1824},
  year={2000}
}

@article{wu2020efficient,
  title={Efficient Sampling-Based Maximum Entropy Inverse Reinforcement Learning with Application to Autonomous Driving},
  author={Wu, Zheng and Sun, Liting and Zhan, Wei and Yang, Chenyu and Tomizuka, Masayoshi},
  journal={IEEE Robotics and Automation Letters},
  volume={5},
  number={4},
  pages={5355--5362},
  year={2020}
}

@inproceedings{lopez2018microscopic,
  title={Microscopic Traffic Simulation using {SUMO}},
  author={Lopez, Pablo Alvarez and Behrisch, Michael and Bieker-Walz, Laura and Erdmann, Jakob and Fl{\"o}tter{\"o}d, Yun-Pang and Hilbrich, Robert and L{\"u}cken, Leonhard and Rummel, Johannes and Wagner, Peter and Wie{\ss}ner, Evamarie},
  booktitle={Proc. 21st IEEE Int. Conf. Intelligent Transportation Systems (ITSC)},
  pages={2575--2582},
  year={2018}
}

@phdthesis{krauss1998microscopic,
  title={Microscopic Modeling of Traffic Flow: Investigation of Collision Free Vehicle Dynamics},
  author={Krau{\ss}, Stefan},
  school={University of Cologne},
  year={1998}
}

@inproceedings{tassa2012synthesis,
  title={Synthesis and stabilization of complex behaviors through online trajectory optimization},
  author={Tassa, Yuval and Erez, Tom and Todorov, Emanuel},
  booktitle={Proc. IEEE/RSJ Int. Conf. Intelligent Robots and Systems (IROS)},
  pages={4906--4913},
  year={2012}
}

@inproceedings{tassa2014control,
  title={Control-limited differential dynamic programming},
  author={Tassa, Yuval and Mansard, Nicolas and Todorov, Emo},
  booktitle={Proc. IEEE Int. Conf. Robotics and Automation (ICRA)},
  pages={1168--1175},
  year={2014}
}

@article{holm1979simple,
  title={A simple sequentially rejective multiple test procedure},
  author={Holm, Sture},
  journal={Scandinavian Journal of Statistics},
  volume={6},
  number={2},
  pages={65--70},
  year={1979}
}

@article{brun2001practical,
  title={Practical identifiability analysis of large environmental simulation models},
  author={Brun, Roland and Reichert, Peter and K{\"u}nsch, Hans R.},
  journal={Water Resources Research},
  volume={37},
  number={4},
  pages={1015--1030},
  year={2001}
}

@article{jones1998efficient,
  title={Efficient global optimization of expensive black-box functions},
  author={Jones, Donald R. and Schonlau, Matthias and Welch, William J.},
  journal={Journal of Global Optimization},
  volume={13},
  number={4},
  pages={455--492},
  year={1998}
}

@inproceedings{ziebart2008maximum,
  title={Maximum entropy inverse reinforcement learning},
  author={Ziebart, Brian D. and Maas, Andrew L. and Bagnell, J. Andrew and Dey, Anind K.},
  booktitle={Proc. AAAI Conf. Artificial Intelligence},
  pages={1433--1438},
  year={2008}
}

@book{nocedal2006numerical,
  title={Numerical Optimization},
  author={Nocedal, Jorge and Wright, Stephen J.},
  edition={2nd},
  publisher={Springer},
  year={2006}
}

@inproceedings{messaoud2021trajectory,
  title={Trajectory Prediction for Autonomous Driving based on Multi-Head Attention with Joint Agent-Map Representation},
  author={Messaoud, Kaouther and Deo, Nachiket and Trivedi, Mohan M. and Nashashibi, Fawzi},
  booktitle={Proc. IEEE Intelligent Vehicles Symposium (IV)},
  pages={165--170},
  year={2021}
}

@inproceedings{liang2020learning,
  title={Learning Lane Graph Representations for Motion Forecasting},
  author={Liang, Ming and Yang, Bin and Hu, Rui and Chen, Yun and Liao, Renjie and Feng, Song and Urtasun, Raquel},
  booktitle={Proc. European Conference on Computer Vision (ECCV)},
  pages={541--556},
  year={2020}
}

@article{kang2017game,
  author  = {Kang, Kyungwon and Rakha, Hesham A.},
  title   = {Game Theoretical Approach to Model Decision Making for Merging Maneuvers at Freeway On-Ramps},
  journal = {Transportation Research Record},
  volume  = {2623},
  number  = {1},
  pages   = {19--28},
  year    = {2017},
  doi     = {10.3141/2623-03}
}

@article{ali2019game,
  author  = {Ali, Yasir and Zheng, Zuduo and Haque, Md. Mazharul and Wang, Meng},
  title   = {A Game Theory-Based Approach for Modelling Mandatory Lane-Changing Behaviour in a Connected Environment},
  journal = {Transportation Research Part C: Emerging Technologies},
  volume  = {106},
  pages   = {220--242},
  year    = {2019},
  doi     = {10.1016/j.trc.2019.07.011}
}

@article{ji2020review,
  author  = {Ji, Ang and Levinson, David},
  title   = {A Review of Game Theory Models of Lane Changing},
  journal = {Transportmetrica A: Transport Science},
  volume  = {16},
  number  = {3},
  pages   = {1628--1647},
  year    = {2020},
  doi     = {10.1080/23249935.2020.1770368}
}

@article{wang2021game,
  author  = {Wang, Mingyu and Wang, Zijian and Talbot, John and Gerdes, J. Christian and Schwager, Mac},
  title   = {Game-Theoretic Planning for Self-Driving Cars in Multivehicle Competitive Scenarios},
  journal = {IEEE Transactions on Robotics},
  volume  = {37},
  number  = {4},
  pages   = {1313--1325},
  year    = {2021},
  doi     = {10.1109/TRO.2020.3047521}
}

@article{laine2023computation,
  author  = {Laine, Forrest and Fridovich-Keil, David and Chiu, Chih-Yuan and Tomlin, Claire},
  title   = {The Computation of Approximate Generalized Feedback {Nash} Equilibria},
  journal = {SIAM Journal on Optimization},
  volume  = {33},
  number  = {1},
  pages   = {294--318},
  year    = {2023},
  doi     = {10.1137/21M142530X}
}

@inproceedings{lidard2024blending,
  author    = {Lidard, Justin and Hu, Haimin and Hancock, Asher and Zhang, Zixu and Contreras, Albert Gim{\'o} and Modi, Vikash and DeCastro, Jonathan and Gopinath, Deepak and Rosman, Guy and Leonard, Naomi Ehrich and Santos, Mar{\'i}a and Fisac, Jaime Fern{\'a}ndez},
  title     = {Blending Data-Driven Priors in Dynamic Games},
  booktitle = {Proceedings of Robotics: Science and Systems XX},
  year      = {2024},
  doi       = {10.15607/RSS.2024.XX.020}
}

@misc{li2026choose,
  author        = {Li, Yueyuan and Nie, Rongcheng and Xi, Weijie and Jiang, Mingyang and Zhang, Songan and Zhuang, Hanyang and Yang, Ming},
  title         = {Choose Your Game Wisely: Measuring Game-Theoretic Structures in Real-World Vehicle Interactions},
  year          = {2026},
  eprint        = {2608.25917},
  archivePrefix = {arXiv},
  primaryClass  = {cs.AI}
}

@inproceedings{peters2021inferring,
  author    = {Peters, Lasse and Fridovich-Keil, David and Rubies-Royo, Vicen{\c{c}} and Tomlin, Claire J. and Stachniss, Cyrill},
  title     = {Inferring Objectives in Continuous Dynamic Games from Noise-Corrupted Partial State Observations},
  booktitle = {Proceedings of Robotics: Science and Systems XVII},
  year      = {2021},
  doi       = {10.15607/RSS.2021.XVII.030}
}

@article{peters2023online,
  author  = {Peters, Lasse and Rubies-Royo, Vicen{\c{c}} and Tomlin, Claire J. and Ferranti, Laura and Alonso-Mora, Javier and Stachniss, Cyrill and Fridovich-Keil, David},
  title   = {Online and Offline Learning of Player Objectives from Partial Observations in Dynamic Games},
  journal = {The International Journal of Robotics Research},
  volume  = {42},
  number  = {10},
  pages   = {917--937},
  year    = {2023},
  doi     = {10.1177/02783649231182453}
}

@inproceedings{li2023cost,
  author    = {Li, Jingqi and Chiu, Chih-Yuan and Peters, Lasse and Sojoudi, Somayeh and Tomlin, Claire and Fridovich-Keil, David},
  title     = {Cost Inference for Feedback Dynamic Games from Noisy Partial State Observations and Incomplete Trajectories},
  booktitle = {Proceedings of the International Conference on Autonomous Agents and Multiagent Systems (AAMAS)},
  pages     = {1062--1070},
  year      = {2023}
}

@article{mehr2023maximum,
  author  = {Mehr, Negar and Wang, Mingyu and Bhatt, Maulik and Schwager, Mac},
  title   = {Maximum-Entropy Multi-Agent Dynamic Games: Forward and Inverse Solutions},
  journal = {IEEE Transactions on Robotics},
  volume  = {39},
  number  = {3},
  pages   = {1801--1815},
  year    = {2023},
  doi     = {10.1109/TRO.2022.3232300}
}

@inproceedings{geiger2021learning,
  author    = {Geiger, Philipp and Straehle, Christoph-Nikolas},
  title     = {Learning Game-Theoretic Models of Multiagent Trajectories Using Implicit Layers},
  booktitle = {Proceedings of the AAAI Conference on Artificial Intelligence},
  volume    = {35},
  number    = {6},
  pages     = {4950--4958},
  year      = {2021},
  doi       = {10.1609/aaai.v35i6.16628}
}

@misc{lian2025game,
  author        = {Lian, Yuansheng and Zhang, Ke and Li, Meng and Li, Shen},
  title         = {Game-Theoretic Modeling of Vehicle Unprotected Left Turns Considering Drivers' Bounded Rationality},
  year          = {2025},
  eprint        = {2507.03002},
  archivePrefix = {arXiv},
  primaryClass  = {cs.AI}
}

@article{treiber2006delays,
  author  = {Treiber, Martin and Kesting, Arne and Helbing, Dirk},
  title   = {Delays, Inaccuracies and Anticipation in Microscopic Traffic Models},
  journal = {Physica A: Statistical Mechanics and its Applications},
  volume  = {360},
  number  = {1},
  pages   = {71--88},
  year    = {2006},
  doi     = {10.1016/j.physa.2005.05.001}
}

@inproceedings{sun2018probabilistic,
  author    = {Sun, Liting and Zhan, Wei and Tomizuka, Masayoshi},
  title     = {Probabilistic Prediction of Interactive Driving Behavior via Hierarchical Inverse Reinforcement Learning},
  booktitle = {Proc. 21st IEEE Int. Conf. Intelligent Transportation Systems (ITSC)},
  pages     = {2111--2117},
  year      = {2018},
  doi       = {10.1109/ITSC.2018.8569453}
}

@article{huang2022driving,
  author  = {Huang, Zhiyu and Wu, Jingda and Lv, Chen},
  title   = {Driving Behavior Modeling Using Naturalistic Human Driving Data with Inverse Reinforcement Learning},
  journal = {IEEE Transactions on Intelligent Transportation Systems},
  volume  = {23},
  number  = {8},
  pages   = {10239--10251},
  year    = {2022},
  doi     = {10.1109/TITS.2021.3088935}
}

@inproceedings{suo2021trafficsim,
  author    = {Suo, Simon and Regalado, Sebastian and Casas, Sergio and Urtasun, Raquel},
  title     = {{TrafficSim}: Learning to Simulate Realistic Multi-Agent Behaviors},
  booktitle = {Proc. IEEE/CVF Conf. Computer Vision and Pattern Recognition (CVPR)},
  pages     = {10400--10409},
  year      = {2021},
  doi       = {10.1109/CVPR46437.2021.01026}
}

@inproceedings{huang2023gameformer,
  author    = {Huang, Zhiyu and Liu, Haochen and Lv, Chen},
  title     = {{GameFormer}: Game-Theoretic Modeling and Learning of Transformer-Based Interactive Prediction and Planning for Autonomous Driving},
  booktitle = {Proc. IEEE/CVF Int. Conf. Computer Vision (ICCV)},
  pages     = {3903--3913},
  year      = {2023}
}

@inproceedings{montali2023waymo,
  author    = {Montali, Nico and Lambert, John and Mougin, Paul and Kuefler, Alex and Rhinehart, Nicholas and Li, Michelle and Gulino, Cole and Emrich, Tristan and Yang, Zoey and Whiteson, Shimon and White, Brandyn and Anguelov, Dragomir},
  title     = {The {Waymo} Open Sim Agents Challenge},
  booktitle = {Advances in Neural Information Processing Systems},
  volume    = {36},
  pages     = {59151--59171},
  year      = {2023}
}

@article{khatib1986real,
  author  = {Khatib, Oussama},
  title   = {Real-Time Obstacle Avoidance for Manipulators and Mobile Robots},
  journal = {The International Journal of Robotics Research},
  volume  = {5},
  number  = {1},
  pages   = {90--98},
  year    = {1986},
  doi     = {10.1177/027836498600500106}
}

@inproceedings{wolf2008artificial,
  author    = {Wolf, Michael T. and Burdick, Joel W.},
  title     = {Artificial Potential Functions for Highway Driving with Collision Avoidance},
  booktitle = {Proc. IEEE Int. Conf. Robotics and Automation (ICRA)},
  pages     = {3731--3736},
  year      = {2008},
  doi       = {10.1109/ROBOT.2008.4543783}
}

@article{gonzalez2016review,
  author  = {Gonz{\'a}lez, David and P{\'e}rez, Joshu{\'e} and Milan{\'e}s, Vicente and Nashashibi, Fawzi},
  title   = {A Review of Motion Planning Techniques for Automated Vehicles},
  journal = {IEEE Transactions on Intelligent Transportation Systems},
  volume  = {17},
  number  = {4},
  pages   = {1135--1145},
  year    = {2016},
  doi     = {10.1109/TITS.2015.2498841}
}

@article{perlin2002improving,
  author  = {Perlin, Ken},
  title   = {Improving Noise},
  journal = {ACM Transactions on Graphics},
  volume  = {21},
  number  = {3},
  pages   = {681--682},
  year    = {2002},
  doi     = {10.1145/566654.566636}
}

@article{cameron2015practitioner,
  author  = {Cameron, A. Colin and Miller, Douglas L.},
  title   = {A Practitioner's Guide to Cluster-Robust Inference},
  journal = {Journal of Human Resources},
  volume  = {50},
  number  = {2},
  pages   = {317--372},
  year    = {2015},
  doi     = {10.3368/jhr.50.2.317}
}

@article{allen1978analysis,
  author  = {Allen, Brian L. and Shin, B. Tom and Cooper, Peter J.},
  title   = {Analysis of Traffic Conflicts and Collisions},
  journal = {Transportation Research Record},
  volume  = {667},
  pages   = {67--74},
  year    = {1978}
}

@article{byrd1995limited,
  author  = {Byrd, Richard H. and Lu, Peihuang and Nocedal, Jorge and Zhu, Ciyou},
  title   = {A Limited Memory Algorithm for Bound Constrained Optimization},
  journal = {SIAM Journal on Scientific Computing},
  volume  = {16},
  number  = {5},
  pages   = {1190--1208},
  year    = {1995},
  doi     = {10.1137/0916069}
}

@inproceedings{wu2024smart,
  author    = {Wu, Wei and Feng, Xiaoxin and Gao, Ziyan and Kan, Yuheng},
  title     = {{SMART}: Scalable Multi-Agent Real-Time Motion Generation via Next-Token Prediction},
  booktitle = {Advances in Neural Information Processing Systems},
  volume    = {37},
  year      = {2024}
}

@misc{hegde2025complete1,
  author       = {Hegde, Sharika and Mahmassani, Hani and Raymer, Meredith and Khakpour, Amirmohammad},
  title        = {Making {CAV} Deployments Compatible with {Complete Streets} Objectives for Safe and Efficient Operation: {Phase I}},
  year         = {2025},
  howpublished = {figshare preprint, doi: 10.6084/m9.figshare.30402241}
}

@misc{hegde2026complete2,
  author       = {Hegde, Sharika and Mahmassani, Hani and Raymer, Meredith and Khakpour, Amirmohammad},
  title        = {Making {CAV} Deployments Compatible with {Complete Streets} Objectives for Safe and Efficient Operation: {Phase II}},
  year         = {2026},
  howpublished = {figshare preprint, doi: 10.6084/m9.figshare.31021405}
}

@misc{hegde2026street,
  author        = {Hegde, Sharika J. and Khakpour, Amirmohammad and Beigi, Pedram and Nie, Yu (Marco) and Mahmassani, Hani S.},
  title         = {{Complete Street} Design Informed by Multi-Agent Simulation},
  year          = {2026},
  eprint        = {2610.27509},
  archivePrefix = {arXiv},
  primaryClass  = {cs.MA}
}

\end{document}